\documentclass[a4paper,11pt]{article}

\usepackage[utf8]{inputenc}
\usepackage[T1]{fontenc}

\usepackage{jcapmod}

\usepackage{amsmath,amsfonts,amssymb}
\usepackage{physics}
\usepackage{mathtools}
\usepackage{bm}

\usepackage{soul}

\usepackage{hyperref}

\usepackage{graphicx}

\usepackage[a4paper,margin=2.5cm]{geometry}

\begin{document}

\pagenumbering{roman}

\begin{titlepage}

\baselineskip=15.5pt \thispagestyle{empty}

\begin{center}
    {\fontsize{19}{24}\selectfont \bfseries  Kalb-Ramond backgrounds in $\alpha'$-complete cosmology}
\end{center}

\vspace{0.1cm}

\begin{center}
    {\fontsize{12}{18}\selectfont Heliudson Bernardo,$^{1}$ Paul R. Chouha$^{1,2}$ and Guilherme Franzmann$^{3}$}
\end{center}

\begin{center}
    \vskip8pt
    \textsl{$^{1}$ Department of Physics, Ernest Rutherford Physics Building, McGill University,\\
        3600 rue Universit\'e, Montr\'eal, Qu\'ebec H3A 2T8, Canada}\\
        
 \textsl{$^{2}$ McGill University School of Continuing Studies,\\
688 Sherbrooke St. West, Montreal, Quebec Canada H3A 3R1}\\
   
    \textsl{$^{3}$ Nordita, KTH Royal Institute of Technology and Stockholm University,\\
        Roslagstullsbacken 23, SE-106 91 Stockholm, Sweden}\\
\end{center}

\vspace{1.2cm}

\hrule
\vspace{0.3cm}

\noindent {\bf Abstract}\\[0.1cm]

We study the matter-coupled equations of motion for cosmological NS massless fields including all $\alpha'$ corrections in an O$(d,d)$ duality invariant approach, with emphasis on the Kalb-Ramond two-form field $B_{(2)}$ and its source. Solutions for the vacuum and matter case are found and the corresponding Einstein frame cosmologies are discussed. We also show that the ansatz for $B_{(2)}$ required by the duality invariant framework implies that the two-form is non-isotropic. 

\end{titlepage}

\thispagestyle{empty}

\setcounter{page}{2}
\tableofcontents
\newpage
\pagenumbering{arabic}
\setcounter{page}{1}
\clearpage




\section{Introduction and Motivation}
\label{sec:intro}

If the effective field theory (EFT) paradigm is correct \cite{Carroll:2021yum}, it is unlikely that we can probe quantum gravity (QG) in its nonperturbative regime.
Moreover, from an EFT perspective, there is little reason to expect that general relativity will break down before we reach energies close to the Planck scale. Although the same hope does not hold for the matter content of the Universe, this is less important if our primary goal is to model the evolution of the Universe as a whole. After all, nothing has prevented us so far from building the successful Standard Model of Cosmology ($\mathrm{\Lambda}$CDM) despite our lack of understanding of the nature of dark matter and dark energy. Thus, in searching for QG effects in cosmology, our attention must be focused on refining our understanding of gravity at high energies where nonperturbative QG plays a significant role. This is one of the underpinnings of the QG program.

We consider string theory to be our main candidate for this program. As we assume that nature's fundamental degrees of freedom are described by strings, we are automatically given another perturbation scale to consider when we insist on recovering our typical point-like picture through field theory.
This scale is parametrized by the string length/scale, $l_\mathrm{s}:= \sqrt{\alpha'}$. Taken together with the string coupling that describes how strongly strings are self-interacting and defines the quantum regime of the theory, these two parameters determine the two perturbative schemes of the theory: length scales well above the string scale that can be described by standard (point-like) field theory, and the weakly coupled regime that can be described by standard quantum field theoretic computations of loop corrections, respectively. 

As expected, there is a regime of the theory in which neither schema is appropriate and where nonpertubative quantum effects are relevant. In this regime, the fundamental ontology of the Universe may not any longer be given by well-defined spacetime quantities. The very notion of spacetime becomes fuzzy due to its quantum nature, and it is not even clear what to make of its Cosmology. As interesting as this can get, we are not attempting to address such a regime here. Thus, throughout this work, we will always assume that we can keep the quantum corrections tamed - the string coupling, modulated by the dilaton field, is weak. Still, much before we get to this convoluted regime, we need to better understand all the string corrections from the classical field theory perspective since perturbations can no longer be trusted  when the energy scale considered is of the order of the string scale. As this regime remains grounded in well-defined geometrical quantities, Cosmology remains a window into its understanding. Surprisingly, though, only recently have we been able to drop its curtains and have a clearer picture of the possible cosmological scenarios that exist in this nonperturbative classical landscape. 

The intuitive picture of seeing strings as particles as we zoom out is still precarious quantitatively. The field theory content of the strings that captures its particle-like behavior has its dynamics fixed by avoiding quantum anomalies in string theory (the so called Weyl anomaly). Unfortunately, to guarantee that the theory is anomaly-free, one has to compute the string corrections order by order, a task that promptly becomes drastically involved. At least, this is how the story goes for general backgrounds. In \cite{Hohm:2019jgu}, Hohm \& Zwiebach found a way to classify all string corrections in a time-dependent background for the massless modes of the strings. They made use of a noncompact symmetry acting on the field space \cite{Sen:1991zi,Maharana:1992my} for those sorts of backgrounds, namely the global O$(d,d)$ symmetry \cite{Meissner:1991ge}. This symmetry generalizes the discrete scale factor duality, which dualizes the physics of a universe having a typical curvature radius, $R$, with that of a universe having the inverse radius $1/R$ (in string units) \cite{Veneziano:1991ek}.

Naturally, it remained to include the infinity tower of massive string modes in the framework. This daunting task can be simplified by considering these modes to define the matter sector of the theory, while the massless modes define its gravitational counterpart. This approach has been first implemented in \cite{Bernardo:2019bkz}, establishing the framework of $\alpha'$-complete cosmology. Many applications to string cosmology have been considered so far \cite{Hohm:2019jgu,Hohm:2019ccp,Krishnan:2019mkv, Wang:2019kez,Wang:2019dcj,Bernardo:2020zlc,Bernardo:2020nol,Bernardo:2020bpa,Quintin:2021eup}, including the construction of a realistic cosmological model \cite{Bernardo:2020nol} which describes the dynamics of a universe that evolves from a phase with topology $\mathbb{R}\times\mathrm{T}^9$ towards four large spacetime dimensions, while keeping the remaining dimensions confined close to the string length, thus making use of the nonperturbative classical regime of the framework. Nonetheless, all those works have assumed that one of the massless degrees of freedom of the string to vanish. 

The universal massless NS-NS sector of the strings is composed of string excitations described by three different fields: the spacetime metric ($G_{\mu\nu})$, the two-form, or Kalb-Ramond (KR) field, ($B_{\mu\nu}$), and dilaton ($\phi$). These fields are spacetime representations of the Lorentz group, but they can be reshuffled to form representations of the O$(d,d)$ group: the determinant of the spatial part of the metric combines with the dilaton field to compose the shifted dilaton ($\Phi$), while the components of the spatial metric and the two-form field combine to form the generalized metric ($S_{MN}$). They transform as a scalar and a rank 2 tensor under the O$(d,d)$ group, respectively. Therefore, O$(d,d)$ rotations mix all the massless modes. Naturally, picking up a basis under this symmetry which fixes any of these fields to be constant breaks this symmetry explicitly if the other fields remain dynamic. Nonetheless, as it was noticed already in the seminal paper \cite{Hohm:2019jgu}, the string corrections were dramatically simplified when the two-form field was set to zero. Thus, the follow-up works generically considered FRLW backgrounds without a flux, paying the price of collapsing the O$(d,d)$ group to the discrete scale factor duality.\footnote{As we comment below, there is one exception by Nunez \textit{et al.} \cite{Nunez:2020hxx}. Their work came out as this paper was being written. While in our work we focus on adding matter to a background that has a non-vanishing two-form field in the particular setting for which the multi-trace terms vanish (or contribute as single-trace ones in the case of an even number of spatial dimensions), their focus was to work out the vacuum equations of motion including the multi-trace terms in generic cosmological backgrounds, and then find interesting solutions. We believe insights from both works will be needed to develop the final equations that will take into account all the $\alpha'$ corrections in an arbitrary background in the presence of matter sources.}

This paper aims to make progress restoring the full O$(d,d)$ symmetry into the $\alpha'$-complete cosmology. The ultimate goal is to have equations of motion (EOM) for any cosmological background including all $\alpha'$ corrections that are O$(d,d)$ invariant in the presence of matter. There has already been one fundamental milestone towards this direction. In \cite{Nunez:2020hxx}, the EOMs in vacuum were derived for anisotropic cosmological backgrounds in the presence of an arbitrary two-form field. In our paper, we consider a more restrictive ansatz which allows us to work with only the simplest set of $\alpha'$ corrections that appear as single-trace contributions of arbitrary powers of $\dot{S}_{MN}^2$  in the action. Under our ansatz, we assume that all the non-vanishing eigenvalues associated with the two-form field are the same ($\beta$), which enhances the remaining symmetry from just having the discrete scale factor duality to also including rotations and reflections in the plane defined by $\beta$ and the Hubble parameter ($H$). In spite of this restrictive ansatz, we show that for an even number of spatial dimensions the multi-trace terms also contribute as single-trace ones, while for an odd number of spatial dimensions the same happens only when the background in the string frame is static ($H=0$). Thus, to keep things general regardless of the number of spatial dimensions, we investigate static solutions both in vacuum and in the presence of matter, possibly paving a new way for the mechanism of volume stabilization by fluxes that is nonperturbative in string corrections and valid up to the string scale. We conclude by studying the corresponding cosmology of these solutions in the Einstein frame, and discuss their feasibility in isotropic spacetimes, as well as their physical relevance for string cosmological models such as string gas cosmology \cite{Brandenberger:1988aj,Bernardo:2020nol}.

\paragraph*{Outline} The paper is organized as follows. In Sec. \ref{Sec 2}, we calculate the components of the single-trace vacuum equations of motion of the O$(d,d)$ invariant action, including a non-trivial KR field. We discuss the conditions for these equations to include all $\alpha'$ corrections and the enhancing of the scale factor duality symmetry by the NS flux. Assuming a duality invariant matter sector, the matter coupled equations are obtained in Sec. \ref{sec:matter_coupling}, where the components of the O$(d,d)$ covariant energy-momentum tensor are calculated, including the current of the $B$-field. Solutions for both the vacuum and matter coupled equations are found in Sec. \ref{Sec 4}, and the corresponding Einstein frame evolution is discussed in Sec. \ref{Sec 5}. We revisit the symmetries of the O$(d,d)$ ansatz for the $B$-field in Sec. \ref{Sec 6}, where we also discuss the relevance of the $B$-source in string gas cosmology. We summarize the results in Sec.~\ref{Sec 7}, together with further discussions.

\section{Single-trace equations of motion in vacuum}\label{Sec 2}

The $\alpha'$-complete cosmology action including matter sources ($S_{\rm m}$) is given by \cite{Hohm:2019jgu,Bernardo:2019bkz}
\begin{equation}\label{theaction}
    S = \frac{1}{2\kappa^2}\int d^dx dt n(t)  e^{-\Phi}\left[-(\mathcal{D}\Phi)^2 - X(\mathcal{DS})\right] + S_{\rm m}\,,
\end{equation}
where a spatial volume integral is omitted, $\kappa^2 \propto \ell_\mathrm{s}^{d-1}$ essentially defines the $D$-dimensional Newton's constant (the proportionality factor varies in accordance with the 10-dimensional theory being considered),
$n(t)$ denotes the lapse function, the shifted dilaton is defined as an $O(d,d)$-invariant scalar field, $\Phi(t) := 2\phi - \frac{1}{2}\ln g$, where $g$ is the determinant of the spacetime metric's spatial part, and the matrix $\mathcal{S}$ is an $O(d,d)$ rank $2$ tensor defined\footnote{The $2d\times2d$ matrices are denoted with calligraphic letters throughout the paper.} as
\begin{align}
    \mathcal{S} :=  \begin{pmatrix}
    bg^{-1} & g- bg^{-1}b \\
    g^{-1} & - g^{-1}b  
    \end{pmatrix} \,,
    \end{align}
constructed out of the matrix components of the two-form field $b_{ij}$, the spatial metric $g_{ij}$, and its inverse.
The arbitrary function $X(\mathcal{DS})$ depends on the first order covariant time-derivative of $\mathcal{S}$, where $\mathcal{D}_t:= n^{-1}\partial_t$.
At lowest order in $\alpha'$, $X(\mathcal{DS}) = (1/8)\text{tr}(\mathcal{DS})^2$, while at higher orders there are both single-trace contributions of arbitrarily high powers of $(\mathcal{DS})^2$ as well as multi-trace ones. In this section, we will write down the single-trace contributions to the equations of motion (EOMs) in vacuum, and then discuss in Sec. \ref{Sec 2.3} the sufficient conditions to be imposed on the cosmological background that will guarantee that the multi-trace terms can safely be ignored, so that the EOM indeed include all $\alpha'$ corrections.

\subsection{Calculating traces with nontrivial flux} \label{sec:derivations} 

From the action \eqref{theaction}, the EOMs including only the single-trace $\alpha'$-corrections in the action are
\begin{subequations}\label{eq:EOM_matter}
    \begin{align}\label{eq:EOM_Phi_matter}
         2\mathcal{D}^2{\Phi} - (\mathcal{D}\Phi)^2 - \sum_{k=1}^{\infty}\alpha'^{k-1}c_k \text{tr}(\mathcal{DS})^{2k} &= \kappa^2 e^{\Phi} \bar{\sigma}\,, \\\label{eq:EOM_n_matter}
        (\mathcal{D}\Phi)^2 - \sum_{k=1}^{\infty}\alpha'^{k-1}(2k-1)c_k \text{tr}(\mathcal{DS})^{2k} &= 2\kappa^2 \Bar{\rho}e^{\Phi}\,, \\\label{eq:EOM_S_matter}
         \mathcal{D}\left[e^{-\Phi}\sum_{k=1}^{\infty}\alpha'^{k-1}4kc_k\mathcal{S}(\mathcal{DS})^{2k-1}\right] &= -\kappa^2 \eta \Bar{\mathcal{T}}\,,
    \end{align}
\end{subequations}
where the dilatonic charge $\sigma$, the energy density $\rho$, and the O$(d,d)$ covariant energy-momentum tensor $\mathcal{T}$ are proportional to variations of the matter action $S_{\rm m}$ with respect to $\Phi$, the lapse $n(t)$, and $\mathcal{S}$, respectively \cite{Bernardo:2019bkz}. The bar over these matter variables denotes multiplication by $\sqrt{g}$, the coefficients $c_k$ mostly remain to be determined from string theory, and $\eta$ is the O$(d,d)$ off-diagonal metric, 
\begin{equation}
\eta = \begin{pmatrix}
    0 & 1 \\
    1 & 0 
    \end{pmatrix} \,.
\end{equation}
These EOMs are only valid for time-dependent backgrounds. Thus, we consider the following ansatz for the fields,
\begin{align}\label{generalansatz}
    ds^2 = -n^2(t)dt^2 + g_{ij}(t)dx^idx^j\,, \quad B_{\mathrm{(2)}} = b_{ij}(t) dx^i\wedge dx^j\,, \quad \phi = \phi(t)\,. 
\end{align}
In order to rewrite the EOMs \eqref{eq:EOM_matter} in terms of the $d$-dimensional fields \eqref{generalansatz} when $b \neq 0$ we need to tackle two issues: i) to write explicitly  $(\mathcal{DS})^{2k}$ and its trace, and ii) to write the O$(d,d)$ covariant energy-momentum tensor $\mathcal{T}$ in terms of the variation of the matter action in terms of $b$ (the current for the KR field cf. Eq. \eqref{currents}). In this section, we show how the first issue is solved while the second one is dealt with in Sec. \ref{sec:matter_coupling}. 

We start off by noticing that 
\begin{equation}
    \mathcal{S} =
   \begin{pmatrix}
    1 & b \\
    0 & 1 
    \end{pmatrix} \begin{pmatrix}
    0 & g \\
    g^{-1} & 0 
    \end{pmatrix} \begin{pmatrix}
    1 & -b \\
    0 & 1 
    \end{pmatrix} := \mathcal{B} \mathcal{S}_0 \mathcal{B}^{-1}\,,
    \end{equation}
where $\mathcal{S}_0$ denotes the matrix $\mathcal{S} $ for $b = 0$. Moreover, $\mathcal{DS}_0 = \mathcal{K}_0 \mathcal{S}_0$, where 
\begin{equation}
    \mathcal{K}_0 \equiv \begin{pmatrix}
    (\mathcal{D}g)g^{-1} & 0 \\
    0 & -g^{-1}(\mathcal{D}g) 
    \end{pmatrix}\,.
\end{equation}
Due to the structure of $\mathcal{B}$ and $\mathcal{DB}$ (both up-triangular matrices, the latter with zeros in the diagonal), the $\mathcal{B}$-matrix has the following properties:
\begin{align}
    (\mathcal{DB}) \mathcal{B} &= \mathcal{DB}\,, \quad \mathcal{B}(\mathcal{DB}) = \mathcal{DB}\,,\quad (\mathcal{DB}) \mathcal{B} ^{-1}= \mathcal{DB}\,, \quad \mathcal{B}^{-1}(\mathcal{DB}) = \mathcal{DB}\,, \quad (\mathcal{DB})^2 = 0\,.
\end{align}
Then, it is straightforward to show 
\begin{equation}
       \{[\mathcal{DB}, \mathcal{S} ], \mathcal{K}
    \mathcal{S} \} =  -\mathcal{B}\{\mathcal{DB}, \mathcal{K}_0\}\mathcal{B}^{-1} -\mathcal{BS}_0\{\mathcal{DB}, \mathcal{K}_0\}\mathcal{S}_0\mathcal{B}^{-1}\,,
\end{equation}
where $\mathcal{K} := \mathcal{B}\mathcal{K}_0\mathcal{B}^{-1}$ and we used the fact that $ \{\mathcal{S} , \mathcal{K}\} = B\{\mathcal{S}_0, \mathcal{K}_0\}B^{-1} =0$. Finally, as we will be computing powers of $\mathcal{DS}$, let us rewrite it as
\begin{equation}\label{eq:DS}
    \mathcal{DS} = [\mathcal{DB}, \mathcal{S}] + \mathcal{K} \mathcal{S} \,,
\end{equation}
which is possible since $\mathcal{DB}^{-1} = - \mathcal{B}^{-1}(\mathcal{DB})\mathcal{B}^{-1}$.
 
To simplify further the calculations, we will assume that $g$ satisfies $\mathcal{D}g \propto g$. This is the case for conformally flat metrics, e.g. a flat FRLW universe in which $g_{ij} = a^2(t) \delta_{ij}$, where $a(t)$ is the scale factor. Then, we have $\{\mathcal{DB}, \mathcal{K}_0\}=0$, and it is easy to show that
\begin{equation}
    \left[[\mathcal{DB}, \mathcal{S} ]^2,  (\mathcal{K}
    \mathcal{S} )^2 \right] =0\,.
\end{equation}
Thus, looking at Eq.~\eqref{eq:DS} we see that $(\mathcal{DS})^{2}$ is a sum of matrices that commute. Therefore, we can use the binomial expansion to rewrite $(\mathcal{DS})^{2k}$ as
\begin{equation}
    (\mathcal{DS})^{2k} = \sum_{j=0}^{k}\binom{k}{j} [\mathcal{DB}, \mathcal{S} ]^{2j}(\mathcal{K}
    \mathcal{S} )^{2(k-j)}\,.
\end{equation}
The expansion can be further simplified by noticing that
\begin{equation}
    [\mathcal{DB}, \mathcal{S} ]^{2j} = [(\mathcal{DB})\mathcal{S}]^{2j} + [\mathcal{S} (\mathcal{DB})]^{2j}\,,
\end{equation}
which follows from showing that $[\mathcal{DB},S_b]^2 = (\mathcal{DB}S_b)^2+(S_b\mathcal{DB})^2$ and then that $(\mathcal{DB}S_b)^2  (S_b \mathcal{DB})^2 \\ = 0$. Thus, we can now rewrite the binomial as
\begin{equation}\label{DS2k}
    (\mathcal{DS})^{2k} = \sum_{j=0}^{k}\binom{k}{j}\left\{[(\mathcal{DB})\mathcal{S}]^{2j}(\mathcal{K}\mathcal{S} )^{2(k-j)} + [\mathcal{S} (\mathcal{DB})]^{2j}(\mathcal{K}\mathcal{S} )^{2(k-j)}\right\}\,.
\end{equation}
Taking the trace of the result above, we can use again $\{\mathcal{S},\mathcal{K}\}=0$ together with the trace's cyclic property to show that
\begin{equation}\label{trS1}
    \text{tr}(\mathcal{DS})^{2k} = 2\sum_{j=0}^{k}\binom{k}{j} \text{tr}\left\{[(\mathcal{DB})\mathcal{S} ]^{2j}(\mathcal{K}\mathcal{S})^{2(k-j)}\right\}\,.
\end{equation}

To write Eq.~\eqref{trS1} in terms of $b(t)$, $a(t)$, and their derivatives, we need to calculate powers of the $2d\times2d$ matrices $(\mathcal{DB})\mathcal{S} $ and $\mathcal{K}\mathcal{S} $. For the former, we have
\begin{equation}
    [(\mathcal{DB})\mathcal{S}_b]^{2j} = \begin{pmatrix}
    [(\mathcal{D}b)g^{-1}]^{2j} & -[(\mathcal{D}b)g^{-1}]^{2j}b \\
    0 & 0 
    \end{pmatrix}\,,
\end{equation}
while for the latter, one can first show that
\begin{equation}
    (\mathcal{K}_b\mathcal{S}_b)^{2l} = (-1)^{l}\mathcal{B}\mathcal{K}_0^{2l}\mathcal{B}^{-1}\,,
\end{equation}
and then that
\begin{equation}\label{KS2l}
    (\mathcal{K}_b\mathcal{S}_b)^{2(k-j)}= (-1)^{(k-j)}\begin{pmatrix}
    [(\mathcal{D}g)g^{-1}]^{2(k-j)} & -[(\mathcal{D}g)g^{-1}]^{2(k-j)}b + b[g^{-1}(\mathcal{D}g)]^{2(k-j)} \\
    0 & [g^{-1}(\mathcal{D}g)]^{2(k-j)} 
    \end{pmatrix}\,.
\end{equation}
Thus, the argument of the trace in \eqref{trS1} can be written as 
\begin{equation}
    [(\mathcal{DB})\mathcal{S}]^{2j}(\mathcal{K}\mathcal{S} )^{2(k-j)} = (-1)^{(k-j)}[(\mathcal{D}b)g^{-1}]^{2j}[(\mathcal{D}g)g^{-1}]^{2(k-j)}\begin{pmatrix}
    1 & -b  \\
    0 & 0
    \end{pmatrix}\,,
\end{equation}
and the sum of $2d$-dimensional traces in Eq.~\eqref{trS1} is reduced to a sum of $d$-dimensional traces:
\begin{equation}
    \text{tr}(\mathcal{DS})^{2k} = 2\sum_{j=0}^{k}\binom{k}{j}(-1)^{(k-j)} \text{tr}\left\{[(\mathcal{D}b)g^{-1}]^{2j}[(\mathcal{D}g)g^{-1}]^{2(k-j)}\right\}\,.
\end{equation}
Finally, writing $\mathcal{D}g = 2H g$, where $H = \mathcal{D}\ln a$ is the Hubble parameter, we get
\begin{align}\label{traceofDS2k}
    \text{tr}(\mathcal{DS})^{2k} &= 2\sum_{j=0}^{k}\binom{k}{j}(-1)^{(k-j)}\text{tr}\left\{[(\mathcal{D}b)g^{-1}]^{2j}\right\}(2H)^{2(k-j)}\,.
\end{align}
From this result, it is clear that \eqref{theaction} depends on $b_{ij}$ through its flux, $\mathcal{D}b_{ij}$, which corresponds to the $0ij$ component of the field strength $H_{\rm (3)} = dB_{\rm (2)}$. This guarantees the invariance of the action under the gauge transformation of the two-form field $B_{\rm (2)}$. Note that for $\mathcal{D}b = 0$, the only term that contributes to the sum is the $j=0$ term, for which we get the same result as in the $b=0$ case. Thus, a constant $b$ does not contribute to the EOMs as expected by gauge symmetry, and we recover the results obtained in \cite{Hohm:2019ccp}. 

\subsection{The O$(d,d)$ charge}

To write the EOMs \eqref{eq:EOM_matter} in terms of $b$ and $H$, we also need to compute the O$(d,d)$ charge operator \cite{Hohm:2019jgu}
\begin{equation}
    \mathcal{Q} = e^{-\Phi}\sum_{k=1}^{\infty}\alpha'^{k-1}4kc_k\mathcal{S}(\mathcal{DS})^{2k-1}\,,
\end{equation}
whose derivative appears on the left-hand side (LHS) of Eq.~\eqref{eq:EOM_S_matter}. In order to proceed, first note that $\mathcal{S}\mathcal{(DS)}^{2k-1} = \mathcal{(DS)}^{2k-2}\mathcal{S}(\mathcal{DS})$. Moreover, it is straightforward to show that
\begin{equation}\label{SDS2}
    \mathcal{S}(\mathcal{DS}) = \begin{pmatrix}
    b g^{-1} (\mathcal{D}b) g^{-1} -(\mathcal{D}g) g^{-1} & -b g^{-1} (\mathcal{D}b) g^{-1} b + (\mathcal{D}g) g^{-1} b + b g^{-1}(\mathcal{D}g) -(\mathcal{D}b) \\
    g^{-1}(\mathcal{D}b) g^{-1} & -g^{-1} (\mathcal{D}b) g^{-1}b + g^{-1} (\mathcal{D}g)
    \end{pmatrix},
\end{equation}
and that
\begin{equation}
    [(\mathcal{DB})\mathcal{S}]^{2j} + [\mathcal{S} (\mathcal{DB})]^{2j} = \begin{pmatrix}
    [(\mathcal{D}b)g^{-1}]^{2j} & -[(\mathcal{D}b)g^{-1}]^{2j}b + b(g^{-1}\mathcal{D}b)^{2j}\\
    0 & (g^{-1}\mathcal{D}b)^{2j}
    \end{pmatrix}.
\end{equation}
Using Eq. \eqref{KS2l} and the last result above into Eq. \eqref{DS2k}, we get
\begin{equation}
    (\mathcal{DS})^{2l} =\sum_{j=0}^{l}\binom{l}{j} (-1)^{l-j}\begin{pmatrix}
    A & B\\
    C & D
    \end{pmatrix},
\end{equation}
where
\begin{align}
    A &= [(\mathcal{D}b) g^{-1}]^{2j}[(\mathcal{D}g) g^{-1}]^{2(l-j)}\,,\nonumber\\
    B &= -[(\mathcal{D}b) g^{-1}]^{2j}\left\{[(\mathcal{D}g)g^{-1}]^{2(l-j)} + b[g^{-1}(\mathcal{D}g)^{2(l-j)}]\right\} +\nonumber\\ &+ \left\{[(\mathcal{D}b)g^{-1}]^{2j} + b [g^{-1}(\mathcal{D}b)]^{2j}\right\}[g^{-1}(\mathcal{D}g)]^{2(l-j)}\,,\nonumber\\
    B &= -[(\mathcal{D}b) g^{-1}]^{2j}[(\mathcal{D}g)g^{-1}]^{2(l-j)}b +b [g^{-1}(\mathcal{D}b)]^{2j}[g^{-1}(\mathcal{D}g)]^{2(l-j)}\,, \nonumber\\
    C &= 0\,,\nonumber\\
    D &= [g^{-1}(\mathcal{D}b)]^{2j}[g^{-1}(\mathcal{D}g)]^{2(l-j)}\,.\nonumber
\end{align}

From this result, using Eq. \eqref{SDS2} and the fact that $\mathcal{D}g \propto g$, we get 
\begin{equation}
    \mathcal{S}(\mathcal{DS})^{2k-1} = \begin{pmatrix}
    \alpha & \beta \\
    \gamma & \delta
    \end{pmatrix},
\end{equation}
where 
\begin{align}
    \alpha &= -2H \left\{[(\mathcal{D}b)g^{-1}]^2 - 4H^2\right\}^{k-1} + b \left\{[g^{-1}(\mathcal{D}b)]^2 - 4H^2\right\}^{k-1}g^{-1}(\mathcal{D}b)g^{-1}\,,\nonumber\\
    \beta &= \left\{[(\mathcal{D}b)g^{-1}]^2 - 4H^2\right\}^{k-1}(2H b - \mathcal{D}b) + b\left\{[g^{-1}(\mathcal{D}b)]^2 - 4H^2\right\}^{k-1}[2H - g^{-1}(\mathcal{D}b)g^{-1}b]\,,\nonumber\\
    \gamma &= \left\{[g^{-1}(\mathcal{D}b)]^2 - 4H^2\right\}^{k-1}[g^{-1}(\mathcal{D}b)g^{-1}]\,,\nonumber\\
    \delta &= \left\{[g^{-1}(\mathcal{D}b)]^2 - 4H^2\right\}^{k-1}[2H-g^{-1}(\mathcal{D}b)g^{-1}b]\,.
\end{align}
This implies that the O$(d,d)$ charge can be written as
\begin{align}\label{oddcharge}
    &e^{\Phi}\mathcal{Q} =
    &\begin{pmatrix}
    -2 H G + b g^{-1}G (\mathcal{D}b) g^{-1} & G (2H b -\mathcal{D}b)+ b g^{-1}Gg[2H - g^{-1}(\mathcal{D}b)g^{-1}b]\\
    g^{-1}G(\mathcal{D}b)g^{-1} & 2H g^{-1}Gg - g^{-1}G(\mathcal{D}b)g^{-1}b
    \end{pmatrix},
\end{align}
where we have defined the matrix $G$ that depends on $b$ and $H$ as
\begin{equation}\label{Gdefinition}
    G(b,H) := \sum_{k=1}^{\infty}\alpha'^{k-1}4k c_k \left\{[(\mathcal{D}b)g^{-1}]^2 - 4H^2\right\}^{k-1}.
\end{equation}

As a consistency check, one can explicitly show that $\mathcal{Q} \in \mathfrak{so}(d,d)$. This can be easily done by showing that $[G(b,H), (\mathcal{D}b)g^{-1}]=0$.

\subsection{Conditions for the single-trace action to include all $\alpha'$-corrections}\label{Sec 2.3}

The EOMs \eqref{eq:EOM_matter} only include contributions from single-trace operators in the action. For an FRLW ansatz with $b=0$, this is sufficient to describe the fully corrected theory, because for this particular ansatz the multi-trace operators have the same form as the single-trace ones \cite{Hohm:2019jgu}, which simply leads to a shift in the coefficients $c_k$. However, when $b\neq0$, the single-trace operators are not guaranteed anymore to have the same form as the multi-trace ones and, therefore, we need to also take into account the multi-trace contributions to the action \cite{Nunez:2020hxx}. 

Since $\mathcal{D}b$ is a real antisymmetric matrix, it follows from the spectral theorem that its eigenvalues are either zero or pure imaginary: for $d$ odd, one eigenvalue is zero and the others come in pairs differing by a phase, $\pm i\beta_j$; for $d$ even, they always come paired and there is no vanishing eigenvalue for a nontrivial matrix. Hence, the eigenvalues of the symmetric matrix $(\mathcal{D}b)^2g^{-1}$ are always negative or zero. Denoting by $-\beta^2_l$ its eigenvalues, the single-trace operators have the form
\begin{equation}\label{eq:single_trace}
    \text{tr}(\mathcal{DS})^{2k} = 2(-1)^k \sum_{m=1}^{d} (\beta^2_m + 4H^2)^{k} \,,
\end{equation}
while the multi-trace operators have the form
\begin{equation}
\prod_{i =1}^j \text{tr}(\mathcal{DS})^{2l_i}=    \prod_{i=1}^j\left[2(-1)^{l_i}\sum_{m=1}^{d} (\beta^2_m + 4H^2)^{l_i}\right] \,.
\end{equation}

Thus, for $d$ even, if all the eigenvalues are equal, $\beta_m=\beta$, then the single and multi-trace operators have the same form, and thus the single-trace equations of motion do indeed include all $\alpha'$ corrections (although the coefficients $c_k$ have to be appropriately shifted)\footnote{This means that one should be able to show that the equations from the full approach of \cite{Nunez:2020hxx} that included all the multi-traces terms reduce to ours for $d$ even.}.  However, for $d$ odd, even if all nonvanishing eigenvalues are equal, there is a monomial term $(4H^2)^{l_i}$ in the sum inside of the product on the multi-trace operator since at least one of the eigenvalues is zero. This term would generate several cross products of the form $(4H^2)^{l_i}(\beta^2_m + 4H^2)^{l_j}$, differing from the structure of the single-trace terms. Explicitly, if $d$ is odd and if we assume that all \emph{nonvanishing} eigenvalues are the same, we have
\begin{align}
    \prod_{i=1}^j\left[2(-1)^{l_i}\sum_{m=1}^{d} (\beta^2_m + 4H^2)^{l_i}\right] = \prod_{i=1}^j\left[2(-1)^{l_i}(d-1) (\beta^2 + 4H^2)^{l_i} + 2(-1)^{l_i}(4H^2)^{l_i}\right], 
\end{align}
and, hence, this will contribute as the single-trace operators only if $H=0$. So, we have arrived at the following conditions for including a nontrivial $b$ in the $\alpha'$-corrected single-trace equations: \textit{for even $d$, it is sufficient that all eigenvalues be the same, while for odd $d$ the only way to include a nonvanishing $B$-field such that the multi-trace terms contribute as single-trace ones is to also assume that $H=0$.} Note that the last possibility also works for $d$ even, but in this case we already have a less restrictive condition.

Note that the condition $\beta_l = 0$ also forces the multi-trace operators to contribute as the single-trace ones, regardless of the number of spatial dimensions. This was originally shown in \cite{Hohm:2019ccp}. In this case, we are back to the ansatz for a flat FRLW metric and vanishing $B$-field. As in the present work we are interested in investigating solutions with nontrivial $b(t)$, we keep $\beta_l\neq0$.

\subsection{Vacuum equations}

Once we have written the trace of $(\mathcal{DS})^{2k}$ and the O$(d,d)$ charge in terms of $b$ and $H$, we can now proceed to find the EOMs in terms of these variables. With the condition $\beta^2_l =\beta^2$ on the eigenvalues of $(\mathcal{D}b)^2g^{-1}$ in mind, we can take the single-trace action to describe our system for $n(t),\, S(t)$ and $\Phi(t)$ with the EOMs given by \eqref{eq:EOM_matter}. 

In this section we focus on the vacuum EOMs, thus turning off the matter sector, $S_m = 0$. The equation for $\mathcal{S}$ becomes
\begin{equation}
    \mathcal{DQ} = 0\,,
\end{equation}
and, upon using Eq. \eqref{oddcharge}, we get four $d \times d$ matrix equations \emph{a priori}. However, the diagonal equations and off-diagonal equations are the same. Explicitly, denoting
\begin{equation}
    \mathcal{DQ} = \mathcal{D}\begin{pmatrix}
    m & n\\
    o & p
    \end{pmatrix}\,,
\end{equation}
the equation for the `$21$' component, $\mathcal{D}o = 0$, is
\begin{equation}\label{bequation1}
    G\mathcal{D}^2 b - (4H +\mathcal{D}\Phi)G \mathcal{D}b + \mathcal{D}G\mathcal{D}b= 0\,,
\end{equation}
and using this equation into $\mathcal{D}m=0$ we get
\begin{equation}\label{Hdotequation}
    (\mathcal{D}b)^2g^{-1}G + 2(H\mathcal{D}\Phi -\mathcal{D}H)G -2H\mathcal{D}G = 0\,.
\end{equation}
Then, using Eq. \eqref{bequation1} and Eq. \eqref{Hdotequation}, one can check that the remaining equations are trivially satisfied.

The Hamiltonian constraint and the equation for $\Phi$ can respectively be written as
\begin{align}
    (\mathcal{D}\Phi)^2 + E(b,H) &= 0\,,\\
    2 \mathcal{D}^2\Phi - (\mathcal{D}\Phi)^2 + F(b,H) &= 0\,,
\end{align}
 where the functions $E(b,H)$ and $F(b,H)$ are defined as
 \begin{align}
    E(b,H) &:= -2 \sum_{k=1}^{\infty}\sum_{l=1}^{d}(-1)^k\alpha'^{k-1}(2k-1)c_k (\beta^2_l + 4H^2)^{k},\label{Edefinition}\\
    F(b,H) &:= -2 \sum_{k=1}^{\infty}\sum_{l=1}^{d}(-1)^k\alpha'^{k-1}c_k (\beta^2_l +4H^2)^{k}\,,\label{Fdefinition}
 \end{align}
 where we have used Eq. \eqref{eq:single_trace} to express the trace of $(\mathcal{DS})^{2k}$ in terms of the eigenvalues of the matrix $(\mathcal{D}b)^2g^{-1}$. Both functions depend on $\beta_l$ and $H$ through the combination $x_l = \beta^2_l + 4H^2$. It is straightforward to show that
 \begin{equation}
     E (x_1,\dots,x_d)= 2\sum_{m=1}^d x_m F_m'(x_m) -F(x_m)\,,
 \end{equation}
 where $F_m$ is the contribution of a single direction to the expression of $F$:
 \begin{equation}
     F_m (x_m):= -2 \sum_{k=1}^{\infty}(-1)^k\alpha'^{k-1}c_k x_m^{k}\,.
 \end{equation}
 
 Summarizing, the single-trace EOMs for the  NS-NS sector in a cosmological background are given by
\begin{subequations}\label{singletraceequations}
\begin{align}\label{neq}
    (\mathcal{D}\Phi)^2 + E  &= 0\,,\\ \label{Phieq}
    2 \mathcal{D}^2\Phi - (\mathcal{D}\Phi)^2 + F &= 0\,,\\
    \label{beq}
    G\mathcal{D}^2 b - (4H +\mathcal{D}\Phi)G \mathcal{D}b + \mathcal{D}G\mathcal{D}b &= 0\,,\\
    \label{aeq}
    (\mathcal{D}b)^2g^{-1}G + 2(H\mathcal{D}\Phi -\mathcal{D}H)G -2H\mathcal{D}G &= 0\,,
\end{align}
\end{subequations}
providing equations of motion for the lapse, $\Phi$, $b(t)$, and $a(t)$, respectively. The functions $E(b,H)$ and $F(b,H)$ are defined above, and the $G(b,H)$ is the $d\times d$ matrix defined in Eq. \eqref{Gdefinition}. Note that we are using the matrix notation $A$ for a covariant 2-tensor $A_{ij}$ such that $(A^2)_{ij} = A_{ik}g^{kl}A_{lj} = (Ag^{-1}A)_{ij}$.
 
As we discussed in Sec. \ref{Sec 2.3}, the sufficient conditions for neglecting multi-trace operators are different for even and odd numbers of spatial dimensions, while we need to have $\beta_l=\beta$ for all nonvanishing eigenvalues despite the number of dimensions. For $d$ even, this implies that
\begin{equation}
    E_\mathrm{{d = 2n}}(b,H) = 2(\beta^2+4H^2) F' - F\,,
\end{equation}
where the prime denotes derivative w.r.t $x$. For $d$ odd, we have a term corresponding to the null eigenvalue, say $\beta_d$, that contains a factor of $x_d = 4H^2$, and only vanishes if we impose that $H=0$. Then, we have
\begin{equation}
    E_\mathrm{{d=2n+1}}(b, 0) = 2\beta^2 F' - F\,.
\end{equation}
We need to consider these differences because they ensure that the solutions to the single-trace EOMs \eqref{singletraceequations} are also solutions to the fully $\alpha'$-corrected theory. Note that if we consider $H=0$ also in the $d$ even case, then the expressions for $E_{d=2n}$ and $E_{d=2n+1}$ have the same structure.  We will specialize to these static backgrounds when we search for solutions to the EOMs in vacuum (see Sec. \ref{sec:vacuum_sol}) and with matter sources (see Sec. \ref{sec:matter_sol}).

\paragraph{A note on symmetries} A closer look at Eqs. \eqref{singletraceequations} and their derivation shows that the discrete symmetry $H\to -H$ of the component equations for $b=0$  studied in \cite{Hohm:2019jgu,Bernardo:2019bkz} is now enhanced to a continuous group together with a discrete symmetry. Since $H$ and $\beta$ always enter in the equations through the combination $x = \beta^2 +H^2$, rotations and reflections in the $(\beta, H)$ plane preserve the form of the equations of motion. The restriction of the general ansatz \eqref{generalansatz} to an FRLW metric and $B$-field such that $\beta_l = \beta$ thus breaks the full O$(d,d)$ symmetry of Eqs. \eqref{eq:EOM_matter} down to the O$(2)$ subgroup. However, for $d$ odd, the symmetry group preserving the solutions will again be discrete, since we need to impose $H=0$ so that the solutions are indeed valid for the full theory, and only the $\beta \to -\beta$ symmetry is left.

\paragraph{Consistency limits} Before studying the matter coupling, we end this section by considering two simple checks. First, we show how to recover the equations discussed previously in the literature \cite{Hohm:2019jgu,Hohm:2019ccp,Krishnan:2019mkv, Wang:2019kez,Wang:2019dcj,Bernardo:2020zlc,Bernardo:2020nol,Bernardo:2020bpa} for a vanishing $B$-field and then we show that we can indeed recover the lowest-order limit ruling the NS-NS sector. 

For the first check, we start by setting $b = 0$. Then, all the $x_l$'s are the same and we have that
 \begin{equation}
     E(0,H) = H\partial_H F(0,H) - F(0,H)\,,
 \end{equation}
 while $G$ is proportional to the identity,
 \begin{equation}
     G_{i}^{\;j}(0,H) = \frac{1}{4dH}\partial_H F(0,H) \delta_{i}^{\;j}\,.
 \end{equation}
 Then, Eqs. \eqref{neq}, \eqref{Phieq}, and \eqref{aeq} reduce to
\begin{align}
    (\mathcal{D}\Phi)^2 + H\partial_H F(0,H) -F(0,H) &= 0\,,\\
    2\mathcal{D}^2\Phi - (\mathcal{D}\Phi)^2 +F(0,H) &= 0\,,\\
    \left[\mathcal{D}\Phi \partial_H F(0,H) - \mathcal{D}( \partial_H F(0,H))\right]\delta_{i}^{\;j} &= 0\,, 
\end{align}
while Eq. \eqref{beq} is identically satisfied. These are the vacuum equations first derived in \cite{Hohm:2019jgu} provided that we identify their $F(H)$ with $F(0,H)$ (which motivated the definitions made in this section).

For the second check, we want to compare the lowest-order limit of Eqs. \eqref{singletraceequations} with the second order equations of motion from the manifestly-diffeomorphism invariant NS-NS action. To do so, we need to truncate $F(b,H)$ and $G(b,H)$ to lowest order in $\alpha'$. In that limit, turning off the $\alpha'$ corrections, there are no multi-trace contributions to the action, and we don't need to impose any restriction on $b(t)$. Using
\begin{equation}
    \sum_{l=1}^d\beta_l^2 = -\text{tr}[(\mathcal{D}b)^2g^{-1}] =- \mathcal{D}b_{ij}\mathcal{D}b^{ji},    
\end{equation}
to evaluate the expressions \eqref{Edefinition} and \eqref{Fdefinition} to lowest-order, we find that
\begin{align}
    E(b,H) &= -2c_1(\mathcal{D}b_{ij}\mathcal{D}b^{ji} - 4dH^2) +\dots\\
    F(b,H) &= -2c_1(\mathcal{D}b_{ij}\mathcal{D}b^{ji} - 4dH^2) +\dots\\
    G(b,H)_i^{\;j} &= 4c_1 \delta_i^{\;j} +\dots \,.
\end{align}
Then, together with $c_1 = -1/8$, the EOMs \eqref{singletraceequations} lead to the lowest-order NS-NS sector equations
\begin{align}\label{lapseeq}
    d(d-1)H^2 + 4\dot{\phi}^2 - 4dH \dot{\phi} +\frac{1}{4}\dot{b}_{ij}\dot{b}^{ji} &= 0\,,\\\label{dilatoneq}
    4\Ddot{\phi} + 4dH\dot{\phi} - 2d\dot{H} - d(d+1)H^2 - 4\dot{\phi}^2 +\frac{1}{4}\dot{b}_{ij}\dot{b}^{ji} &= 0\,,\\\label{bfieldeq}
    \Ddot{b}_{ij} - (4H +2 \dot{\phi} -dH)\dot{b}_{ij} &=0\,,\\\label{scalefactoreq}
    \dot{b}_{il}\dot{b}^{l}_{\;\;j} + 2 g_{ij}(2H\dot{\phi} -\dot{H} -dH^2) &=0\,,
\end{align}
where we assumed $n(t) =1$ and used $\dot{\Phi} = 2\dot{\phi} - dH$ (the overdots denote standard time-derivative). These are exactly the EOMs coming from the action
\begin{equation}\label{NSaction}
    S = \frac{1}{2\kappa^2} \int d^Dx \sqrt{-G}e^{-2\phi}\left[R + 4(\partial\phi)^2 -\frac{1}{2}|H_{\rm (3)}|^2 \right]
\end{equation}
once we impose the ansatz \eqref{generalansatz}. Note that $H_{\rm (3)}$ is the flux associated with the $B$-field (cf. Eq. \eqref{eq:flux}). See Appendix \ref{appendixA} for more details.

\section{The matter-coupled equations} \label{sec:matter_coupling}

To get the component equations in terms of $H(t)$ and $b(t)$ including contributions from a general O$(d,d)$ invariant matter action $S_{\text{m}}$ we need to compute the components of the O$(d,d)$ covariant energy-momentum tensor $\mathcal{T}$. As it has been shown in \cite{Bernardo:2019bkz} the $\alpha'$ corrections do not change the lowest-order matter coupling after field redefinitions, thus the components of $\mathcal{T}$ keep the same form before and after the infinite tower of $\alpha'$ corrections are considered. Keeping that in mind, in this section we write down the components of $\mathcal{T}$. 

Starting from Eq. \eqref{eq:EOM_S_matter}, we write its LHS and RHS as the following matrices
\begin{equation}\label{DQequation}
    \mathcal{DQ} = \mathcal{D}\begin{pmatrix}
    m & n\\
    o & p
    \end{pmatrix} = -\kappa^2 \begin{pmatrix}
    \alpha & \beta\\
    \gamma & \delta
    \end{pmatrix}\,,
\end{equation}
such that their components give off four $d\times d$ equations:
\begin{align}
    \mathcal{D}o &= e^{-\Phi}g^{-1}\left[G \mathcal{D}^2b + \mathcal{D}G \mathcal{D}b - (4H + \mathcal{D}\Phi)G \mathcal{D}b\right]g^{-1} = -\kappa^2 \gamma \,,\\
    \mathcal{D}m &= e^{-\Phi}\left\{bg^{-1}\left[G \mathcal{D}^2b + \mathcal{D}G \mathcal{D}b - (4H + \mathcal{D}\Phi)G \mathcal{D}b\right]g^{-1} +\right. \nonumber\\
    &+ \left. (\mathcal{D}b)^2g^{-1}G + 2(H\mathcal{D}\Phi- \mathcal{D}H)G-2H \mathcal{D}G\right\} = -\kappa^2 \alpha \,,\\
    \mathcal{D}n &= e^{-\Phi}\left\{-bg^{-1}\left[G \mathcal{D}^2b + \mathcal{D}G \mathcal{D}b - (4H + \mathcal{D}\Phi)G \mathcal{D}b\right]g^{-1}b - \right. \nonumber\\
    &-\left. \left[G \mathcal{D}^2b + \mathcal{D}G \mathcal{D}b - (4H + \mathcal{D}\Phi)G \mathcal{D}b\right] - \right.\nonumber\\
    &- \left. \left[(\mathcal{D}b)^2g^{-1}G + 2(H\mathcal{D}\Phi- \mathcal{D}H)G-2H \mathcal{D}G\right]b - \right.\nonumber \\
    &-\left. bg^{-1}\left[(\mathcal{D}b)^2g^{-1}G + 2(H\mathcal{D}\Phi- \mathcal{D}H)G-2H \mathcal{D}G\right]g   \right\} = -\kappa^2 \beta \,,\\
    \mathcal{D}p &= e^{-\Phi}\left\{-g^{-1}\left[G \mathcal{D}^2b + \mathcal{D}G \mathcal{D}b - (4H + \mathcal{D}\Phi)G \mathcal{D}b\right]g^{-1}b - \right. \nonumber\\
    &- \left. g^{-1}\left[(\mathcal{D}b)^2g^{-1}G + 2(H\mathcal{D}\Phi- \mathcal{D}H)G-2H \mathcal{D}G\right]g \right\} = -\kappa^2 \delta\,.
\end{align}
Using the first equation into the second, we get
\begin{equation}\label{correctedb2equationmatter}
    e^{-\Phi}\left[(\mathcal{D}b)^2g^{-1}G + 2(H\mathcal{D}\Phi- \mathcal{D}H)G-2H \mathcal{D}G\right] = -\kappa^2(\alpha - b\gamma)\,,
\end{equation}
and using this result together with the first equation in the expression for $n$ and for $p$, we have that
\begin{align}
    \beta &= -b\gamma b - g\gamma g - (\alpha -b\gamma) b - bg^{-1}(\alpha -b \gamma)g\,, \\
    \delta &= -\gamma b - g^{-1}(\alpha -b \gamma)g\,.
\end{align}

Then, if we find $\alpha$ and $\gamma$ in terms of variations of the matter action, we can write $\eta \bar{\mathcal{T}}$ in terms of these variations. To do that, we can readily use the lowest-order results in Appendix \ref{appendixA}. In fact, plugging Eq. \eqref{NSTijeq} into Eq. \eqref{NSsigmaeq}, we find that
\begin{equation}\label{b2equationmatter}
    \dot{b}_{il}\dot{b}^{l}_{\;\;j} + 2 g_{ij}(2H\dot{\phi} -\dot{H} -dH^2) = -2\kappa^2e^{2\phi}\left(T_{ij}^{\text{t}} -\frac{\sigma_{\phi}}{4}g_{ij}\right) \equiv -2\kappa^2 e^{2\phi} T_{ij}\,,
\end{equation}
where we denoted the energy-momentum tensor at constant $\Phi$ by $T_{ij}$. Note that this is a natural definition, since given an O$(d,d)$ invariant matter action
\begin{equation}
    S_{\text{m}}[n, g, \phi] = S_{\text{m}}[n, \mathcal{S}, \Phi]\,,
\end{equation}
the total spatial energy-momentum tensor $T^{\text{t}}_{ij}$ includes contributions from the metric dependence on $\Phi$, while if we vary the action with respect to $\mathcal{S}$ with $\Phi$ fixed, only the metric dependence coming from $\mathcal{S}$ will contribute. Hence, the O$(d,d)$ covariant energy-momentum tensor $\mathcal{T}$ will have $T_{ij}$ among its components, and not $T^{\text{t}}_{ij}$. See \cite{Quintin:2021eup} for a detailed discussion. In the following, note that the index-suppressed $T$ stands for $T_{i}^{\;j}$, while $J$ corresponds to $J^{ij}$. Recall also that the bar over the variables denotes multiplication by $\sqrt{g}$.

Comparing Eq. \eqref{b2equationmatter} with the lowest-order limit of Eq. \eqref{correctedb2equationmatter} (recall that $G_{i}^{\;j}(b,H) =  4c_1\delta_{i}^{\;j} = -(1/2)\delta_{i}^{\;j}$), we have that
\begin{equation}
    \alpha -b\gamma = - \bar{T}\,.
\end{equation}
Similarly, comparing the `$ij$' component of the lowest-order equation for $B_{\mu\nu}$ (Eq. \eqref{NSJijeq})
\begin{equation}
    \Ddot{b}_{ij} - (4H -dH + 2\dot{\phi})\dot{b}_{ij} = - 2\kappa^2 e^{2\phi}J_{ij}\,,
\end{equation}
with the equation for the `$o$' component, we read off that
\begin{equation}
    \gamma = -\bar{J}\,.
\end{equation}
Therefore, we finally find that
\begin{align}
    \alpha &= -\bar{T} - b\bar{J}\,,\\
    \beta &= b\bar{J} b + g\bar{J} g +\bar{T}b + bg^{-1}\bar{T}g\,,\\
    \gamma &= -\bar{J}\,,\\
    \delta &= \bar{J}b + g^{-1}\bar{T}g\,.
\end{align}
That is, comparing Eq. \eqref{DQequation} with Eq. \eqref{eq:EOM_S_matter}, what we have shown is that, even after considering the whole tower of $\alpha'$ corrections, the O$(d,d)$ covariant energy-momentum tensor still reads as
\begin{equation}
    \bar{\mathcal{T}} =
    \begin{pmatrix}
    -\bar{J} & \bar{J}b + g^{-1}\bar{T}g\\
    -\bar{T} - b\bar{J} & b\bar{J} b + g\bar{J} g +\bar{T}b + bg^{-1}\bar{T}g
    \end{pmatrix}\,,
\end{equation}
which has the same form as the O$(d,d)$ energy-momentum tensor introduced in \cite{Gasperini:1991ak}, modulo  some sign differences due to the differences in conventions. As discussed above, this was expected since in \cite{Bernardo:2019bkz} it was shown that the matter sector's description remains unchanged, whether one starts with the corrected
equations and then sources them with matter, or one considers corrections to the matter-sourced
lowest-order equations.

Now, we can summarize the components of the single-trace EOMs including matter to all orders in $\alpha'$: 
\begin{subequations} \label{matter_equations}
\begin{align}
    (\mathcal{D}\Phi)^2 + E &= 2\kappa^2 e^{\Phi}\Bar{\rho}\,,\\ 
    2 \mathcal{D}^2\Phi - (\mathcal{D}\Phi)^2 + F &= \kappa^2 e^{\Phi} \bar{\sigma}\,,\\
    G\mathcal{D}^2 b - (4H +\mathcal{D}\Phi)G \mathcal{D}b + \mathcal{D}G\mathcal{D}b &= \kappa^2 e^{\Phi}g\bar{J}g\,,\\
    (\mathcal{D}b)^2g^{-1}G + 2(H\mathcal{D}\Phi -\mathcal{D}H)G -2H\mathcal{D}G &= \kappa^2 e^{\Phi}\bar{T}\,.
\end{align}
\end{subequations}
For completeness, we write down the O$(d,d)$ continuity equation as well 
\begin{equation}
    \mathcal{D}\bar{\rho} + \frac{1}{4}\text{tr}(\mathcal{SDS}\eta \bar{\mathcal{T}}) -\frac{1}{2}\mathcal{D}\Phi\bar{\sigma} = 0\,.
\end{equation}
In terms of $H$ and $\mathcal{D}b$, it is straightforward to show that it can be written as
\begin{equation}
\mathcal{D}\bar{\rho} + dH\bar{p}+ \frac{1}{2}\text{tr}[(\mathcal{D}b)\bar{J}]-\frac{1}{2}\mathcal{D}\Phi\bar{\sigma} = 0\,.
\end{equation}

\section{Solutions to the Single-Trace Equations of Motion}\label{Sec 4}

Now that we have derived the single-trace $\alpha'$-corrected EOMs coupled with matter sources, we can proceed to look for solutions. As it was emphasized in Sec. \ref{Sec 2.3}, these equations encapsulate all the string corrections as long as we consider static solutions in the string frame, i.e. $H=0$ (although this is not needed for even number of spatial dimensions), and that all the eigenvalues of the $B$-field are the same. We first consider a vacuum background and then a few cases including matter sources. 

\subsection{Vacuum solutions} \label{sec:vacuum_sol}

For the vacuum case, we can focus on Eqs.  \eqref{singletraceequations}. We impose the condition $H=0$ and simplify the equations by considering $\Phi = 2\phi$ and $n(t) =1$, such that $\mathcal{D}$ reduces to the standard time-derivative. The metric is constant and taken to be Minkowski's, though for notation consistency we write the spatial metric $g$ explicitly. Thus, the equations reduce to 
\begin{subequations}
    \begin{align}
    4\dot{\phi}^2 - \frac{2}{[d]}\text{tr}(\dot{b}^2 g^{-1})F' - F  &= 0\,,\\ 
    4 \ddot{\phi} - 4\dot{\phi}^2 + F &= 0\,,\\
    G \ddot{b} - 2 \dot{\phi} G \dot{b} + \dot{G} \dot{b} &= 0\,,\\
    \dot{b}^2 g^{-1}G &= 0\,,
    \end{align}    
\end{subequations}
where $[d]=d$  ($[d]=d-1$) when $d$ is even (odd). From the definition of $G_i^{\;j}$ in Eq. \eqref{Gdefinition}, the basis that diagonalizes  $\dot{b}^2g^{-1}$ also diagonalizes $G$. So, we can easily take the trace of the last equation to get
\begin{equation}
    -2 \beta^2 F' = 0 \implies F'= 0\,, 
\end{equation}
and recall that $F'$ was defined as the derivative of $F$ with respect to the variable $x = \beta^2 +4H^2$, which reduces to $\beta^2$ when $H=0$. In the diagonal basis of $G$, we can easily check that 
\begin{equation}
    G= \frac{2}{[d]}F'\text{diag}(1,\cdots,1,0)\,, \label{diagonal_G}
\end{equation}
where the null entry in the diagonal only appears when $d$ is odd, since in this case there is one vanishing eigenvalue of $\text{tr}(\dot{b}^2g^{-1})$, as explained in Sec. \ref{Sec 2.3}. Hence, $F'= 0$ implies $G = 0$ (one could also see this by simply realizing that because we are assuming $\beta^2\neq0$ the last equation of motion implies $G = 0$). Then, the EOM for $b$ is trivially satisfied. Moreover, the constraint equation gives
\begin{equation}
    4\dot{\phi}^2 = F(\beta)\,,
\end{equation}
and $\dot{\phi}$ should be constant as the condition $F'=0$ only makes sense if $\beta$ is constant. The EOM for $\phi$ is consistently satisfied. Thus, we find the solution 
\begin{equation}\label{Phi_dot_cst}
    H = 0\,,\qquad \dot{\phi}^2 = \frac{F(\beta)}{4}\,, \qquad \dot{b} g^{-1} = \beta \text{diag}(i\sigma_2, i\sigma_2\,, \dots,i\sigma_2, 0)\,,
\end{equation}
where there is an even number of Pauli matrices $\sigma_2$ in  $(\mathcal{D}b)$, that is, the null matrix in the last entry is only there if $d$ is odd. This solution has a linear dilaton whose velocity is fixed by the flux, and is only a solution in string theory if the function $F'$ has a nontrivial root. Note that it is a nonperturbative solution: if we expand $F$ and keep only the first term, then $F'= 0$ implies $\beta= 0$ and $F(\beta=0) =0$, so we get only the trivial Minkowski solution with constant dilaton and $B_{\mu\nu} = 0$.

This simple solution is less trivial than one would imagine. Looking back at Eq. \eqref{generalansatz}, we have initially assumed a homogeneous ansatz for all the fields. However, throughout the derivation of the equations in Sec. \ref{sec:derivations}, we also assumed that the metric was isotropic. The above solution, nonetheless, presents a solution for the $b$ field which is anisotropic. This is only possible nonperturbatively, as the perturbative solution in vacuum necessarily fixes $\beta=0$. We will discuss these matters thoroughly in Sec. \ref{sec:anisotropies}.

\subsection{Matter-sourced solutions} \label{sec:matter_sol}

Similarly to the vacuum case above, we start from Eq. \eqref{matter_equations}, and using Eq. \eqref{diagonal_G} - which is valid regardless of the matter sources - our working equations are
\begin{subequations}\label{mattereq}
\begin{align}\label{mattereqforn}
   4 \dot{\phi}^2 - \frac{2}{[d]}\text{tr}(\dot{b}^2 g^{-1})F' - F  &= 2\kappa^2 e^{2\phi}\rho\,,\\ \label{mattereqforphi}
    4 \ddot{\phi} - 4\dot{\phi}^2 + F &= \kappa^2 e^{2\phi} \sigma\,,\\ \label{mattereqforb}
    G\ddot{b} -2\dot{\phi}G\dot{b} + \dot{G}\dot{b} &= \kappa^2 e^{2\phi}gJg\,,\\ \label{mattereqforg}
    \dot{b}^2g^{-1}G &= \kappa^2 e^{2\phi}T\,,
\end{align}
\end{subequations}
where we have dropped the barred variables on the right-hand side given that the volume is fixed. Meanwhile, the continuity equation reads 
\begin{equation}
\dot{\rho} + \frac{1}{2}\text{tr}(\dot{b}J) - \dot{\phi} \sigma = 0\,.
\end{equation}

We can further simplify the equations. The spatial energy-momentum tensor for a perfect fluid source is given by $T_{i}^{\;j} = p\delta_i^{\;j}$, where $p$ is the pressure. For a fluid with a barotropic equation of state, the pressure is related to the energy density $\rho$ via $p=w \rho$. Hence, taking the trace of Eq. \eqref{mattereqforg}, we find that
\begin{equation}
    \text{tr}(G\dot{b}^2g^{-1}) = -2 F' \beta^2 =  d\kappa^2 e^{2\phi} p\,,
\end{equation}
which can be solved for $F'$ assuming $\beta \neq 0$, and then substituted in Eq. \eqref{mattereqforn}. Note that for $d$ odd the LHS of Eq. \eqref{mattereqforg} has one vanishing diagonal component (cf. the basis that diagonalizes $\dot{b}^2g^{-1}$), and since orthogonal transformations preserves the form of $T$, the pressure should therefore vanish. Thus, for an odd number of spatial dimensions only presureless matter sources are allowed in a background with nontrivial flux. We will discuss this case separately in the following.

Additionally, we can write Eq. \eqref{mattereqforb} as
\begin{equation}
    \frac{d}{dt}\left(e^{-2\phi}G \dot{b}\right) = \kappa^2 g J g\,,
\end{equation}
which, after multiplication by $\dot{b}g^{-1}$ from the left, by $g^{-1}$ from the right, and then taking the trace, gives
\begin{equation}
    \text{tr}\left[\dot{b}g^{-1}    \frac{d}{dt}\left(e^{-2\phi}G \dot{b}g^{-1}\right)\right] = \kappa^2 \text{tr}(\dot{b}J)\,.
\end{equation}

To further simplify the equations, we consider the constant flux case, $\ddot{b}=0$, which implies that $G$ is also constant. In this case, the eigenvalues $\beta$, the functions $F(\beta)$ and $F'(\beta)$ are also constant, and thus we denote them by $F_\beta$ and $F'_\beta$ in what follows to emphasize that they are constants. Thus, the Eqs. \eqref{mattereq} imply that
\begin{subequations}\label{mattereqwithH=0_&_consteigenvalues}
\begin{align}
    4 \dot{\phi}^2 - F_\beta  &= 2\kappa^2 e^{2\phi} \left(\rho + \frac{d}{2}p\right)\,,\\ 
    4 \ddot{\phi} - 4\dot{\phi}^2 + F_\beta &= \kappa^2 e^{2\phi} \sigma\,,\\
     \text{tr}(\dot{b} J) &=  -2dp \dot{\phi}\,,\label{trace_matter_eq_const_eigen}\\
     2F'_\beta\beta^2 &= -d \kappa^2 e^{2\phi} \rho w\,.
\end{align}
\end{subequations}
Using the third equation, the continuity equation yields
\begin{equation}\label{rho-eq}
    \dot{\rho} - d \dot{\phi}p -\dot{\phi}\sigma = 0\implies \rho =\rho_0 e^{\phi(dw + \lambda)}\,,
\end{equation}
assuming both $w$ and the dilatonic equation of state $\lambda = \sigma/\rho$ to be constants. Plugging this back into the last equation in \eqref{mattereqwithH=0_&_consteigenvalues}, we get
\begin{equation}
    2F'_\beta\beta^2  =- d \kappa^2  w \rho_0 e^{\phi (dw+\lambda+2)}\,.  \label{F_prime_equation}
\end{equation}
Thus, as the LHS is a constant, we can either have that both $F'_\beta = 0$ and $w=0$, or $\rho_0$ vanishing (vacuum case) or that $dw+\lambda+2=0$ for $F'_\beta \neq 0$. While if we plug \eqref{rho-eq} in Eq. \eqref{trace_matter_eq_const_eigen}, we get
\begin{equation}
     \text{tr}(\dot{b}J)  = -2d \dot{\phi}w \rho_0 e^{\phi(dw+\lambda)}\,. \label{gauge_source_equation}
\end{equation}
Finally, adding the first two equations, and using the solution for the energy density, the dilaton's equation reads
\begin{equation}
    4 \ddot{\phi} = \rho_0 \kappa^2 (2+dw+\lambda) e^{\phi(2+dw+\lambda)} \label{dilaton_eq_const_eigen} \,,
\end{equation}
which gives a linear solution when $2+dw+\lambda=0$ and a logarithmic solution when $w=0$, as we will discuss below.\\ 

\subsubsection{$2+dw+\lambda=0$ case}
From Eq. \eqref{dilaton_eq_const_eigen}, $\ddot{\phi}=0$, therefore the dilaton evolves linearly in time, $\phi(t) = \phi_0 + \dot{\phi}_0t$. The solution to the continuity equation now reads 
\begin{align}
    \rho = \rho_0 e^{-2\phi}\,,
\end{align}
and the dilaton's equation in Eq. \eqref{mattereqwithH=0_&_consteigenvalues} thus gives
\begin{align} 
    - 4\dot{\phi}_0^2 + F_\beta = \kappa^2e^{2\phi}\lambda\rho = \kappa^2 \lambda \rho_0 \,,
\end{align}
constraining the value of $F_\beta$, and therefore the eigenvalues $\beta$, together with $\dot{\phi}_0$ and $\lambda$. Meanwhile, Eq. \eqref{F_prime_equation} fixes $F'_\beta$,
\begin{equation}
    2F'_\beta\beta^2 =- d \kappa^2 w \rho_0\,,
\end{equation}
and Eq. \eqref{gauge_source_equation} fixes the evolution of $J$, which decays away following the same evolution as the energy density,
\begin{equation}
     \text{tr}(\dot{b}J)  = -2d \dot{\phi}_0w \rho_0 e^{-2\phi}.
\end{equation}

Note that for a gas of strings, we have $\lambda=0$ (as it can be easily seen from the Polyakov action for a single string cf. Eq. \eqref{eq:Polyakov_action}), and hence the equation of state is given by $w=-2/d$. Thus, this case is not particularly useful for modeling a gas of strings. Nonetheless, we consider the corresponding cosmology in the Einstein frame in Sec. \ref{sec:linear_dil_EF}.

\subsubsection{$F'_\beta = 0 = p$ case}

When $p=0$, the LHS of Eq. \eqref{mattereqforb} implies that for $\dot b\neq 0$, $G=0$.  Using Eq. \eqref{trace_matter_eq_const_eigen}, we thus see that $J=0$. Summing the first two equations, we can find a solution for the dilaton using Eq. \eqref{dilaton_eq_const_eigen}, 
\begin{equation}\label{log-Phi}
    \phi(t) = \phi_0 - \frac{2}{\lambda+2}\ln \left|\frac{t}{t_0}\right|\,.
\end{equation}
This is also not particularly interesting to describe a gas of strings, since in this case we expect  that $J\neq0$ around the self-dual point, which is the point where the equation of state vanishes (see discussion in Sec. \ref{sec:b-field}). Still, the cosmology associated with this solution in the Einstein frame is discussed in Sec. \ref{sec:log_dil_EF}.

\subsubsection{$J=0$ case}

When the $B$-field does not couple with matter, we note from the third equation \eqref{mattereqforb} that the dilaton must be a constant (otherwise $F'_\beta=0$ and from the last equation we have $w=0$, i.e. we are back to the previous case) which also implies that the energy density is a constant. Thus,
\begin{equation}
    F_\beta = \kappa^2 e^{2\phi_0} \sigma_0 = - 2\rho \kappa^2 e^{2\phi_0} \left(1 + \frac{dw_0}{2}\right), \quad \quad F'_\beta \beta^2 = \frac{d}{2} \kappa^2 e^{2\phi_0}\rho_0 w_0\,.
\end{equation}
This describes static solutions in both frames for any constant equation of state supported by the dilatonic charge and the gauge field background. Thus, this solution could also possibly be used to sustain the loitering backgrounds demanded by string gas cosmology \cite{Brandenberger:1988aj,Bernardo:2020nol}.

\paragraph{$\mathbf{\sigma=0.}$} A particular case is when the dilatonic charge is zero. Then, we have $F_\beta = 0$ and $w=-2/d$, while the last equality sets a condition over $F'_\beta$. This also corresponds to a static solution in the Einstein frame completely sustained by the gauge field background with a fixed equation of state.

\section{Einstein frame cosmology}\label{Sec 5}

Although the solutions in the previous sections have $H = 0$, they have a nontrivial dilaton evolution, thus their background evolutions are not trivial in the Einstein frame (EF). Indeed, in making contact with cosmological observations, we would like to study the background evolution in the EF as all the standard cosmology is based on solutions of the EF equations. In this section, after a brief overview of the EF formulae, we consider two types of solutions for the dilaton (the linear and logarithmic evolutions obtained in the previous sections), derive their corresponding cosmologies in the EF, and compare them with the solutions of the $D$-dimensional Friedmann's equations. Note that this sort of analysis was considered in  \cite{Hohm:2019jgu,Bernardo:2019bkz,Quintin:2021eup} as well. 

\subsection{From the string frame to the Einstein frame}

The metric $G^{\mathrm{E}}_{\mu\nu}$ in the EF is related to the string frame (SF) metric $g_{\mu\nu}$ by a Weyl rescaling,
\begin{equation}
    G^{\mathrm{E}}_{\mu\nu}=e^{-4\phi/(d-1)}g_{\mu\nu} \,,
\end{equation}
such that the line element in the EF is given by
\begin{equation}
    ds^2_\mathrm{E}=-dT^2+a_{\mathrm{E}}(T)^2dx_idx^i \,,
\end{equation}
with $T$ denoting the time parameter and $a_{\mathrm{E}}(T)$ the scale factor, both with respect to the EF. In the SF we have that
\begin{equation}
ds^2_\mathrm{S}=-dt^2+a(t)^2dx_idx^i\,.
\end{equation}
To describe the transformations from the SF to the EF, we simply need the time-time and space-space components separately of the Weyl rescaling given that both metrics depend only on the time parameter. They read, respectively,
\begin{subequations}
\begin{align}
    dT &=e^{-2\phi(t)/(d-1)} dt\,, \label{Time-EF} \\
    a_{\mathrm{E}}(T) &=e^{-2\phi(t)/(d-1)}a(t)\,.
\end{align}
\end{subequations}

With these transformations in hand, we can compute the EF Hubble parameter $H_{\mathrm{E}}$,
\begin{equation}\label{Scale-factor-EF}
    H_{\mathrm{E}}(T)\equiv \frac{d a_{\mathrm{E}}(T)}{dT}\frac{1}{a_{\mathrm{E}}(T)} = \left[-\frac{2}{(d-1)}\frac{d\phi(T)}{dT}e^{-2\phi(t)/(d-1)}+H\right]e^{2\phi(t)/(d-1)}  \,.
\end{equation}
Since we are most interested in the $H=0$ case, the expression for the Hubble parameter in the EF thus reduces to
\begin{equation}\label{Hubble-Einstein}
      H_{\mathrm{E}}(T)=-\frac{2}{(d-1)}\frac{d\phi(T)}{dT} \,.
\end{equation}
We can also use the above expression to directly relate the scale factor $a_{\mathrm{E}}(T)$ in the EF with the dilaton using Eq. \eqref{Scale-factor-EF}, which leads upon integration to 
  \begin{equation}\label{Scale-factor-EF-Phi}
    a_{\mathrm{E}}(T)=a_{\mathrm{E}}(T_0)\ e^{-2(\phi-\phi_0)/(d-1)} \,.
\end{equation}

\subsection{Linear Dilaton} \label{sec:linear_dil_EF}
 
According to Eq. \eqref{Phi_dot_cst}, the dilaton in the SF is linear, and thus can be written as
\begin{equation}\label{Phi_lin}
  \phi(t)=\alpha t+\gamma \,.
\end{equation}
Comparing the two expressions in Eq. \eqref{Phi_dot_cst} and in Eq. \eqref{Phi_lin}, we see that $\alpha=\pm\sqrt{F(\beta)}/2$, and therefore $\alpha$ can be either positive or negative. In order to relate the time parameter in both frames, we use Eq. \eqref{Phi_lin} in Eq. \eqref{Time-EF}, and write
\begin{equation}
      dT=e^{-2(\alpha t+\gamma)/(d-1)}dt\,.
\end{equation}
Upon integration, we can read after some algebra that
\begin{equation}\label{time-in-SF-in-terms-of EF}
      t=-\frac{1}{\alpha}\left(\gamma+\frac{d-1}{2}\ln \left|e^{-2\phi_0/(d-1)}-\frac{2\alpha (T-T_0)}{d-1}\right|\right)\,,
\end{equation}
where we have set $\phi(t_0)\equiv\phi_0=\alpha t_0+\gamma$. Therefore, the dilaton as a function of the Einstein-time frame reads,
\begin{equation}\label{Dilaton-in-terms-EF-time}
    \phi(T)=-\frac{d-1}{2}\ln \left|e^{-2\phi_0/(d-1)}-\frac{2\alpha (T-T_0)}{d-1}\right|\,.
\end{equation} 
We can now compute the Hubble parameter in the Einstein frame using Eq. \eqref{Hubble-Einstein},
\begin{equation}\label{Hubble-in-terms-EF-time}
      H_{\mathrm{E}}(T)=\frac{-2\alpha}{(d-1)e^{-2\phi_0/(d-1)}-2\alpha (T-T_0)}\,.
\end{equation} 
For $\alpha<0$, we can adjust the origin of the time coordinate, $T_0$, such that $T$ runs over $(0, \infty)$ for $t$ varying in $(-\infty, \infty)$. In this case, we have that
\begin{equation}
    \phi(T) = -\frac{d-1}{2}\ln\left( \frac{2|\alpha|}{d-1}T\right), \quad  H_{\mathrm{E}}(T) = \frac{1}{T}\,,    
\end{equation}
which corresponds to a scale factor $a_{\mathrm{E}}(T) \propto T$. For comparison, a solution to the $D$-dimensional Friedmann equations with such scale factor corresponds to an effective equation of state $w = (2-d)/d$, and so for $d=3$ the solution is the same we would obtain from the Friedmann equations with a winding equation of state. This solution saturates the strong energy condition (see \cite{Quintin:2021eup}).

\subsection{Logarithmic Dilaton} \label{sec:log_dil_EF}

As done above, we relate the time parameter in both frames by using Eq. \eqref{log-Phi} in Eq. \eqref{Time-EF}. Upon integration, we get
 \begin{equation}\label{Time-EF-Log-Phi}
    T-T_0=\frac{e^{-2\phi_0/(d-1)}}{(n+1)}t_0\left[\left(\frac{t}{t_0}\right)^{n+1}-1\right], \quad \text{where} \quad n\equiv\frac{4}{(\lambda+2)(d-1)}\,.\\
 \end{equation}
 Inverting this relation and following the same procedure showed explicitly for the linear case, we can write the dilaton and the Hubble parameters in the EF as
 \begin{equation}
    \phi(T) = \phi_0 - \frac{2(d-1)}{4 + (\lambda+2)(d-1)}\ln \left|\frac{T}{T_0}\right|\,, \quad H_{\mathrm{E}}(T) = \frac{4}{4 +(\lambda+2)(d-1)}\frac{1}{T}\,,
 \end{equation}
 after picking a convenient choice for the EF time origin. This corresponds to the EF scale factor evolving as $a_{\mathrm{E}}(T) \propto T^{\frac{4}{4 +(\lambda+2)(d-1)}}$. For matter with vanishing dilatonic charge, $\lambda=0$, such as for a string gas source, we find
 \begin{equation}
     a_{\mathrm{E}}(T) \propto T^{\frac{2}{d+1}}\,,
 \end{equation}
 which has the form of a solution to the $D$-dimensional Friedmann equations with an effective radiation equation of state, $w = \frac{1}{d}$. There is also a singularity at $T=0$ both in the metric and the dilaton, which is also present in the SF frame, since the dilaton is singular at $t=0$ (equivalent to $T=0$). We comment on how this solution can be related to the final regime of a string gas based cosmology in the discussion of Sec. \ref{sec:b-field}.
 
 \section{Discussions}\label{Sec 6}

In this section, we comment on some physical aspects arising from our analysis of the $B$-field in $\alpha'$-complete cosmology. We will focus on two particularities: i) the fact that the nonperturbative equations allow cosmological solutions for the $B$-field which are not isotropic despite the metric being rotationally invariant, and ii) the role played by the sources associated with the $B$-field in the context of string gas cosmology.\footnote{The role of the $\Phi$-charge is discussed extensively in \cite{Quintin:2021eup} where it is shown that its presence is vital to recover the Einstein gravity limit from the $\alpha'$-complete cosmology framework.}

\subsection{General analysis of the ansatz for $B_{\mathrm{(2)}}$: conditions for isotropy} \label{sec:anisotropies}

A cosmological background is typically homogeneous and isotropic. This is implemented by considering the Lie derivative of the metric to be zero, $\mathcal{L}_{\xi}g_{\mu\nu} = 0$, where $\xi$ denotes the generators of spatial translations and rotations. In our framework, we also have a time-dependent (i.e., translation invariant) two-form field (cf. \eqref{generalansatz}) with nontrivial solutions (as illustrated in \eqref{Phi_dot_cst}, for example). Is this ansatz also invariant under the rotation group SO$(d)$? 

The general condition for isotropy is
\begin{equation}
    \mathcal{L}_{\xi_{[mn]}}B_{\mu\nu} = \xi_{[mn]}^\rho \partial_\rho B_{\mu\nu} + B_{\rho\nu}\partial_\mu \xi^\rho_{[mn]} + B_{\mu\rho}\partial_\nu \xi^\rho_{[mn]}=0 \,, \label{eq:iso_cond_1}
\end{equation}
where $\xi_{[mn]}$ denotes the $d(d-1)/2$ generators of the $d$-dimensional rotation group. In the space of fields, they take the following representation in Cartesian comoving coordinates
\begin{equation}
    \xi_{[mn]} = \xi^\mu_{[mn]}\partial_\mu = x^m\partial_n - x^n\partial_m \,. \label{eq:rep_generators}
\end{equation}
The $[mn]$ appearing on the LHS is labeling the different numbers of generators (note that $\xi_{[mn]}=-\xi_{[nm]}$, therefore we can consider $m>n$ without loss of generality), while the indices appearing on the RHS are spatial indices. Thus, using Eq.~\eqref{eq:rep_generators} in Eq.~\eqref{eq:iso_cond_1}, the two-form is isotropic if
\begin{align}\label{isocondition}
    \xi^k_{[mn]} \partial_k B_{ij} + \delta_i^m B_{nj} -\delta_i^n B_{mj} + \delta_j^m B_{i n} - \delta_j^n B_{i m} =0\,,
\end{align}
resulting into $d(d-1)/2$ independent equations for each component of $B_{ij}$. Note that the gauge symmetries of $B_{\mu\nu}$ can always be used to put $B_{0i} = 0$, i.e., we need only to consider its spatial components. Now, we can separate them into different cases: $i =m$ and $j\neq n$, $(i,j) = (m,n)$, and $(i,j)\neq \{m,n\}$ such that the conditions for isotropy become, respectively, 
\begin{subequations}
\begin{align}\label{oneindexequal}
   \xi^k_{[mn]} \partial_k B_{mj} + B_{nj} &= 0\,,\\ \label{twoindicesequal}
    \xi^k_{[mn]} \partial_k B_{mn} &= 0\,,\\ \label{twoindicesdifferent}
    \xi^k_{[mn]} \partial_k B_{ij} - \delta^n_i B_{mj} + \delta^m_j B_{i n} &= 0\,,
\end{align}
\end{subequations}
where we also considered $i > j$ without loss of generality as $B_{ij}$ is antisymmetric, and there is no summation over $m$ and $n$. The last equation is only present when $d>3$ since for $d=3$ at least one of the indices of $B_{ij}$ will be equal to $m$ or $n$. For $d>3$, there is no general solution for the conditions above and one needs to consider case by case, e.g. for $d=3$ there is a class of solutions $B_{ij} = (h/3!) \epsilon_{ijk}x^k$ which gives constant flux proportional to the volume form of the spatial slices, $H_{ijk} = h \epsilon_{ijk}$. 

Let us now specialize to the explicit O$(d,d)$ ansatz
\begin{equation}
    B_{\mathrm{(2)}} = b_{ij}(t)dx^i\wedge dx^j\,.
\end{equation}
The general condition for isotropy, Eq. \eqref{isocondition}, gives
\begin{equation}
    \mathcal{L}_{\xi_{[mn]}}b_{ij} = \delta_i^m b_{n j} -\delta_i^n b_{m j} + \delta_j^m b_{i n} - \delta_j^n b_{i m} = 0 \,.
\end{equation}
When $(i,j) = (m,n)$ and $(i,j) \neq \{m,n\}$, this condition is trivially satisfied. However, for $i = m,\, j\neq n$, we get
\begin{equation}
    \mathcal{L}_{\xi_{[mn]}}b_{mj} = b_{n j} = 0\,,
\end{equation}
implying that the only possible isotropic solution is the trivial one with a vanishing $B$-field. We see that the ansatz for the $B_{\mathrm{(2)}}$ compatible with the O$(d,d)$ symmetry (for $d>2$) forces us to consider an anisotropic $B$-field.\footnote{For $d=2$, we would necessarily have $(i,j)=(m,n)$, and only Eq. \eqref{twoindicesequal} would be present. In this case, the O$(1,1)$ ansatz is isotropic.}

A very similar analysis can be done for the flux, $H_{\mu\nu\rho}$, which is the physical gauge invariant quantity. In fact, for a general manifold there is no global description of $B_{\rm (2)}$, i.e., one cannot solve the Bianchi identity $dH_{(3)} = 0$ by writing $H_{(3)} = dB_{\rm (2)}$ globally. This is the case whenever the manifold admits harmonic 2-forms (that is, the second cohomology class $H_2(M, \mathbb{R})$ is nontrivial). So, it is useful to impose the isotropy conditions directly on $H_{(3)}$. For our ansatz, 
\begin{equation}
    H_{\rm (3)} = \dot{b}_{ij}(t)dt\wedge dx^i \wedge dx^j\,,
\end{equation}
and so the only nonzero components are $H_{0ij}(t)$. After some algebra, it is easy to show that isotropy leads to 
\begin{equation}
    \mathcal{L}_{\xi_{[mn]}} H_{m i 0} = H_{n i 0}=0\,,
\end{equation}
which implies that the only isotropic solutions are the ones with vanishing flux. Thus, we arrive at the same result as before: the ansatz for $B_{\mathrm{(2)}}$ does not allow for isotropic solutions, regardless of the number of spatial dimensions. Finally, it is worth mentioning that this analysis could also be done using the Hodge dual of $H_{\rm (3)}$, which defines the axion field in $d=3$, where a homogeneous solution for the axion can only be implemented through a spatial dependent $B$-field \cite{Copeland:1994km}. 

The results above beg the following question: ``how does one obtain solutions with an isotropic metric and nontrivial anisotropic $B_{\mathrm{(2)}}$ field in $\alpha'$-complete cosmology?'' For the lowest-order theory, Einstein's equation imply that the $B$-field is isotropic whenever the metric is isotropic, and if we stick to the O$(d,d)$ ansatz this will result into a vanishing $B$-field. Hence, there could not be perturbative solutions with an isotropic metric and an anisotropic field. However, after the inclusion of the infinite tower of corrections, it seems that the nonlinear theory for $B_{\mathrm{(2)}}$ is such that the total anisotropic contribution for the total energy-momentum tensor vanishes. In this case, we can have a nonperturbative solution with FRLW metric but with an anisotropic flux. 

\subsection{The $B$-field's current from string matter} \label{sec:b-field}

Given that we need to consider the $B$-field together with the metric and the dilaton in order to preserve the O$(d,d)$  symmetry group for the universal massless sector of string theory in cosmological backgrounds, by the same token, we also need to consider $B$-currents, $J_{\mu\nu}$, derived from its coupling to the matter sector if we are to preserve this symmetry across the matter sector. In this section, we explore heuristically the role that this current plays in the presence of a gas of strings.\footnote{Note that this has also been explored in the context of the O$(D,D)$ symmetry group in Double Field Theory \cite{Angus:2018mep,Angus:2019bqs}.}

We start off by considering the spacetime Polyakov action of a single string in a generic background defined by $G_{\mu\nu}$ and $B_{\mu\nu}$,
\begin{equation}\label{eq:Polyakov_action}
    S = -\frac{1}{4\pi \alpha'} \int d^Dx \int d^2\sigma \delta^{(D)}(x - X(\sigma))\left[\sqrt{-\gamma}\gamma^{ab}\partial_a X^\mu \partial_b X^\nu G_{\mu\nu}(X) + \epsilon^{ab}\partial_a X^\mu \partial_b X^\nu B_{\mu\nu}(X) \right]\,.
\end{equation}
Assuming the conformal gauge\footnote{That is the reason why we do not consider the coupling of the string to the dilaton background - it vanishes in this gauge.} throughout this section, a string in a $d$-dimensional torus background with same radii $R$ in all directions and vanishing $B$-field has as solution to its wave EOM
\begin{equation}
    X^\mu(\tau, \sigma) = x^{\mu}_0 +\alpha' p^\mu \tau + w^\mu R \sigma + i \sqrt{\alpha'}\sum_{n\neq 0}\frac{e^{in\tau}}{n}\left(\alpha_n^\mu e^{-in\sigma} + \Tilde{\alpha}^\mu_n e^{in\sigma}\right)\,,
\end{equation}
where $p^\mu = (E, n^i/R)$ is the center of mass momentum of the string and $w^\mu = (0, w^i)$ encodes the winding numbers in any direction. Both $n_i$ and $w^i$ are integers. 

Using Eq. \eqref{currents}, the on-shell energy-momentum tensor and the $B$-current for this action are given by, 
\begin{subequations}
\begin{align}
    T_{\mu\nu} &= - \frac{1}{2\pi \alpha'\sqrt{-G}}\int d\tau d\sigma \left(-\alpha' p_\mu p_\nu + R^2 w_\mu w_\nu \right)\delta (x-X(\sigma))\,, \\
    J^{\mu\nu} &=  \frac{1}{2\pi \alpha'\sqrt{-G}}\int d\tau d\sigma \left(2\alpha'R p^{[\mu}w^{\nu]}\right)\delta (x- X(\sigma))\,,
\end{align}
\end{subequations}
respectively. The results are valid for a single string in a flat background. To go from this to a gas of free strings, we assume a spatial averaging procedure such that the momentum $p^\mu$ and winding $w^\mu$ of each string in the gas contribute to give a homogeneous and isotropic energy-momentum tensor and current. Then, their components are, respectively,
\begin{subequations}
\begin{align}
    T_{\mu\nu} &\leftrightarrow \alpha'\langle p_\mu p_\nu \rangle - R^2 \langle w_\mu w_\nu\rangle\,, \\
    J^{\mu\nu} &\leftrightarrow 2\alpha'R \langle p^{[\mu}w^{\nu]}\rangle\,.
\end{align}
\end{subequations}
Then, considering  $p^2 = -M^2$ and defining the number of left and right oscillatory modes as 
\begin{equation}
    N = \sum_{n \neq 0} \alpha^\mu_n \alpha_{n \mu} \,, \quad \Tilde{N} = \sum_{n\neq0} \Tilde{\alpha}^\mu_n \Tilde{\alpha}_{\mu n}\,,
\end{equation}
we have
\begin{subequations}
\begin{align}
    T_{00} &\sim \alpha'\langle p_0p_0\rangle =  \alpha'\langle M^2 \rangle = \frac{\alpha'}{R^2}\langle n_i n^i\rangle + \frac{R^2}{\alpha'}\langle w_i w^i  \rangle + 2\left(\langle N \rangle + \langle \Tilde{N}\rangle - 2\right)\,, \\
    T_{ij} &\sim \alpha' \langle p_ip_j\rangle - R^2 \langle w_i w_j\rangle\,.
\end{align}
\end{subequations}

Now, assuming the gas of strings to be isotropic, we consider
\begin{equation}
    \langle p_i p_j \rangle = \frac{\langle p_kp^k \rangle}{d}g_{ij} = \frac{\langle n_k n^k\rangle}{R^2 d}g_{ij}, \quad \langle w_i w_j\rangle = \frac{\langle w_k w^k \rangle}{d}g_{ij}\,,
\end{equation}
such that
\begin{equation}
    T_{ij} \sim \frac{\alpha'}{R^2d} \langle n_i n^i\rangle g_{ij} - \frac{R^2}{d}\langle w_i w^i \rangle g_{ij}\,.
\end{equation}
Thus, for a gas dominated by momentum (winding) modes, we have that $p = \rho/d$ ($p = -\rho/d$), while a gas dominated by oscillatory modes has zero pressure since $T_{ij}$ does not depend on $N$ and $\Tilde{N}$, as expected \cite{Battefeld:2005av}. Hence, we have reproduced the analysis of the thermodynamic canonical ensemble.

From the expression of the current, we see that 
\begin{subequations}\label{eq:current_comp}
\begin{align}
    J^{0i} &\sim 2\alpha' R \langle E w^i\rangle\,,\\
    J^{ij} &\sim 2\alpha' \langle n^{[i}w^{j]}\rangle\,.
\end{align}
\end{subequations}
The first equation tell us that in order for the gas to have zero net $B$-charge, the total winding number in each direction should vanish, implying that the winding and antiwinding numbers should cancel each other out. The second equation implies that $J^{ij}$ is negligible for a gas dominated solely by winding or by momentum modes. These results motivate us to consider a nontrivial $J$ only for radii that are close to the string length since apart from this value the gas will be dominated only by one kind of these modes. Note that in string gas cosmology the oscillatory modes dominated phase is expected to be short, so an interesting matter sector ansatz would be to consider a pressureless fluid with constant $J$. 

Therefore, we can finally better understand the role that the $B$-field and its current can play in string gas cosmology. From Eqs. \eqref{eq:current_comp}, we see that the $B$-field sourced by the string gas matter modulates the transition between winding and momentum modes. This is crucial to model the cosmology of a gas of closed strings and should be taken into account when building
realistic early universe cosmological models. Moreover, as emphasized above, this coupling should be mostly relevant around the self-dual point where both winding and momentum modes co-exist. Finally, we also see that we need to go beyond the solutions studied in our paper, as we saw in Sec. \ref{sec:matter_sol} most of the cases considered involved either $J=0$ or a constant flux.

\section{Conclusion}\label{Sec 7}

In this paper, we have studied the matter-coupled equations of motion including the NS 2-form field in $\alpha'$-complete cosmology. We have written down the components of the single-trace equations of motion and found sufficient conditions for them to include the entire tower of $\alpha'$ curvature corrections relevant for cosmological backgrounds. Under some assumptions on $b(t)$ and that of a static metric, the contributions of the multi-trace operators appearing in the duality invariant action are similar in form to those of the single-trace, and hence can be consistently neglected. 

We also found vacuum and matter-coupled solutions with nontrivial $b(t)$. They are nonperturbative solutions to the fully $\alpha'$-corrected duality covariant equations, supported by a dynamical $b(t)$. However, it remains to be checked whether they are genuine string theory backgrounds, since they should satisfy $F'(\beta) = 0$ to exist, and the specific analytical expression of $F$ (which should follow from requiring the quantum worldsheet theory to be conformally invariant) is currently unknown (although see \cite{Basile:2021amb} for an attempt to compute $F(H)$ using functional renormalization group techniques). In \cite{Bonezzi:2021mub} a duality invariant approach to Weyl anomaly cancellation on the worldsheet theory was initiated, and it seems to be a promising avenue to get perturbative information about $F$. On the other hand, as discussed in \cite{Nunez:2020hxx}, nonperturbative information is also necessary. Already in \cite{Alexandre:2006xh,Alexandre:2007dp,Alexandre:2007hs}, non-perturbative time-dependent backgrounds were discussed, where the vanishing of the worldsheet beta functionals were argued for inductively after field redefinitions, and the existence of a non-perturbative infrared fixed point in the wordsheet theory was shown to exist. We leave investigating how these results constrain the functional form of $F$ for future work.   

Although the solutions found have a Minkowski metric in the string frame, the dilaton's evolution is nontrivial, and hence they correspond to nontrivial cosmologies in the Einsten frame. Potential applications to the string gas cosmology scenario were discussed and we have shown that a dynamical $\dot{b}$ is crucial for describing the transition from winding to momentum modes dominated phases. Moreover, due to the ansatz of $b(t)$ required by the O$(d,d)$ symmetry, we have shown that solutions with nontrivial $b(t)$ would necessarily be anisotropic and, for an odd number of spatial dimensions (which is the relevant case in superstring theory), the only consistent equation of state for $H=0$ is that of pressureless matter. 

The nonperturbative solutions with $H=0$ and nontrivial $B$-field lead us to speculate about the possibility of volume stabilization by fluxes. In the type II case, fluxes can be used to stabilize the axion-dilaton and complex structure moduli, while providing a potential with a flat direction for the K\"ahler ones  \cite{Dasgupta:1999ss,Gukov:1999ya,Giddings:2001yu} (see \cite{Grana:2005jc, Blumenhagen:2006ci} for reviews), and the volume moduli require nonperturbative effects to be stabilized \cite{Kachru:2003aw, Balasubramanian:2005zx}. The $\alpha'$ corrections implies a nonlinear theory for $b(t)$, and so it could be that its flux alone could stabilize the volume moduli. However, further investigation is necessary since a full stabilization mechanism would require us to consider a much more general metric than the one considered in the duality invariant setup? (see \cite{Dasgupta:2019gcd, Brahma:2020tak}, and references therein for a discussion on time-dependent fluxes in moduli stabilization).

Finally, as explained in the introduction, the vacuum equations developed in \cite{Nunez:2020hxx} are capable of handling a more general ansatz than the one considered in the present paper. It would be interesting to develop them further by including their coupling with a matter sector. The resulting equations could then be used to find nonperturbative anisotropic solutions with nontrivial $b(t)$ and $g(t)$. We leave this possibility for future work. 

\section*{Acknowledgments}

We thank Robert Brandenberger for his involvement in the early stages of this work and for valuable discussions and insightful comments that shaped many aspects of this work. HB is partially supported by funds from NSERC.

\appendix

\section{The lowest-order theory}\label{appendixA}

THe NS-NS action coupled with matter is given by
\begin{equation}\label{NSactionmatter}
    S = \frac{1}{2\kappa^2} \int d^Dx \sqrt{-G}e^{-2\phi}\left[R + 4(\partial\phi)^2 -\frac{1}{2}|H_{3}|^2 \right] + S_{m}\,.
\end{equation}
The EOMs for the metric $G_{\mu\nu}(x)$, dilaton $\phi(x)$ and Kalb-Ramond field $B_{\mu\nu}(x)$ are, respectively,
\begin{subequations}\label{NSequations}
    \begin{align}\label{metricequation}
        R_{\mu\nu} -\frac{1}{2}G_{\mu\nu}R + 2 \nabla_\mu \nabla_\nu \phi - \frac{1}{4}H_{\mu\alpha\beta}H_{\nu}^{\;\;\alpha\beta} +\frac{1}{2}G_{\mu\nu}\left(4\nabla_\rho \phi \nabla^{\rho}\phi - 4 \nabla^2\phi + \frac{1}{2}|H_{3}|^2\right) &= \kappa^2 e^{2\phi}T_{\mu\nu}^{\text{t}}\,,\\\label{dilatonequation}
    R + 4 \nabla^2\phi - 4\nabla_\rho \phi \nabla^\rho \phi - \frac{1}{2}|H_{3}|^2 &= -\kappa^2 e^{2\phi}\frac{\sigma_{\phi}}{2}\,,\\\label{bfieldequation}
    -\frac{1}{2}\nabla^{\rho}H_{\rho\mu\nu} + (\nabla^{\rho} \phi)H_{\rho\mu\nu} &= 2\kappa^2 e^{2\phi}J_{\mu\nu}\,,
    \end{align}
\end{subequations}
where 
\begin{equation} \label{currents}
    T_{\mu\nu}^{\text{t}} =- \frac{2}{\sqrt{-G}}\frac{\delta S_{m}}{\delta G^{\mu\nu}}\,, \quad 
    \sigma_{\phi} = - \frac{2}{\sqrt{-G}}\frac{\delta S_{m}}{\delta \phi}\,, \quad J^{\mu\nu} = - \frac{2}{\sqrt{-G}}\frac{\delta S_{m}}{\delta B_{\mu\nu}}\,,
\end{equation}
and we are using the notation
\begin{equation}\label{eq:flux}
    \int H_{3} \wedge \ast H_{3} = \int |H_{3}|^2 \sqrt{-G} d^Dx, \quad |H_{3}|^2 = \frac{1}{3!}G^{\mu_1\mu_2}G^{\nu_1\nu_2}G^{\rho_1\rho_2}H_{\mu_1\nu_1\rho_1}H_{\mu_2\nu_2\rho_2} \,,
\end{equation}
with $H_{\rho\mu\nu} = 3\partial_{[\rho}B_{\mu\nu]}$ being the usual field strength for $B_{\mu\nu}$.

Imposing the ansatz (\ref{generalansatz}), Eqs. \eqref{NSequations} reduce to
\begin{subequations}
\begin{align}\label{NSrhoeq}
    d(d-1)H^2 + 4\dot{\phi}^2 - 4dH\dot{\phi} -\frac{1}{4}\dot{b}_{ij}\dot{b}^{ij} &= 2\kappa^2e^{2\phi}\rho\,,\\\label{NSTijeq}
   g_{ij}\left[-(d-1)\dot{H}-\frac{d}{2}(d-1)H^2-2\dot{\phi}^2 + 2\Ddot{\phi}+ 2(d-1)H\dot\phi -\frac{1}{8}\dot{b}_{ij}\dot{b}^{ij}\right] - \frac{1}{2}\dot{b}_{il}\dot{b}^{l}_{\;\;j} &=\kappa^2 e^{2\phi}T_{ij}^{\text{t}}\,,\\\label{NSsigmaeq}
    2d\dot{H} +d(d+1)H^2 - 4(\Ddot{\phi}+dH\dot{\phi}) + 4\dot{\phi}^2 +\frac{1}{4}\dot{b}_{ij}\dot{b}^{ij} &= -\kappa^2 e^{2\phi}\frac{\sigma_{\phi}}{2}\,,\\\label{NSJijeq}
    \Ddot{b}_{ij} - (4H -dH + 2\dot{\phi})\dot{b}_{ij} &= - 2\kappa^2 e^{2\phi}J_{ij}\,.
\end{align}
\end{subequations}

\newpage 
\phantomsection
\addcontentsline{toc}{section}{References}

\let\oldbibliography\thebibliography
\renewcommand{\thebibliography}[1]{
  \oldbibliography{#1}
  \setlength{\itemsep}{0pt} 
  \footnotesize 
}

\bibliographystyle{JHEP2}
\bibliography{References}

\providecommand{\url}[1]{#1}\providecommand{\href}[2]{#2}\begingroup\raggedright\begin{thebibliography}{10}

\bibitem{Carroll:2021yum}
S.M.~Carroll, \emph{{The Quantum Field Theory on Which the Everyday World
  Supervenes}},  \href{https://arxiv.org/abs/2101.07884}{{\ttfamily
  arXiv:2101.07884}}.

\bibitem{Hohm:2019jgu}
O.~Hohm and B.~Zwiebach, \emph{{Duality invariant cosmology to all orders in
  $\alpha$'}}, \href{https://doi.org/10.1103/PhysRevD.100.126011}{\emph{Phys.
  Rev. D} {\bfseries 100} (2019) 126011}
  [\href{https://arxiv.org/abs/1905.06963}{{\ttfamily arXiv:1905.06963}}].

\bibitem{Sen:1991zi}
A.~Sen, \emph{{O(d) x O(d) symmetry of the space of cosmological solutions in
  string theory, scale factor duality and two-dimensional black holes}},
  \href{https://doi.org/10.1016/0370-2693(91)90090-D}{\emph{Phys. Lett. B}
  {\bfseries 271} (1991) 295}.

\bibitem{Maharana:1992my}
J.~Maharana and J.H.~Schwarz, \emph{{Noncompact symmetries in string theory}},
  \href{https://doi.org/10.1016/0550-3213(93)90387-5}{\emph{Nucl. Phys. B}
  {\bfseries 390} (1993) 3}
  [\href{https://arxiv.org/abs/hep-th/9207016}{{\ttfamily hep-th/9207016}}].

\bibitem{Meissner:1991ge}
K.A.~Meissner and G.~Veneziano, \emph{{Manifestly O(d,d) invariant approach to
  space-time dependent string vacua}},
  \href{https://doi.org/10.1142/S0217732391003924}{\emph{Mod. Phys. Lett. A}
  {\bfseries 6} (1991) 3397}
  [\href{https://arxiv.org/abs/hep-th/9110004}{{\ttfamily hep-th/9110004}}].

\bibitem{Veneziano:1991ek}
G.~Veneziano, \emph{{Scale factor duality for classical and quantum strings}},
  \href{https://doi.org/10.1016/0370-2693(91)90055-U}{\emph{Phys. Lett. B}
  {\bfseries 265} (1991) 287}.

\bibitem{Bernardo:2019bkz}
H.~Bernardo, R.~Brandenberger and G.~Franzmann, \emph{{O$(d,d)$ covariant
  string cosmology to all orders in $\alpha^{\prime}$}},
  \href{https://doi.org/10.1007/JHEP02(2020)178}{\emph{JHEP} {\bfseries 02}
  (2020) 178} [\href{https://arxiv.org/abs/1911.00088}{{\ttfamily
  arXiv:1911.00088}}].

\bibitem{Hohm:2019ccp}
O.~Hohm and B.~Zwiebach, \emph{{Non-perturbative de Sitter vacua via $\alpha'$
  corrections}}, \href{https://doi.org/10.1142/S0218271819430028}{\emph{Int. J.
  Mod. Phys. D} {\bfseries 28} (2019) 1943002}
  [\href{https://arxiv.org/abs/1905.06583}{{\ttfamily arXiv:1905.06583}}].

\bibitem{Krishnan:2019mkv}
C.~Krishnan, \emph{{de Sitter, $\alpha'$-Corrections \textbackslash{}\& Duality
  Invariant Cosmology}},
  \href{https://doi.org/10.1088/1475-7516/2019/10/009}{\emph{JCAP} {\bfseries
  10} (2019) 009} [\href{https://arxiv.org/abs/1906.09257}{{\ttfamily
  arXiv:1906.09257}}].

\bibitem{Wang:2019kez}
P.~Wang, H.~Wu, H.~Yang and S.~Ying, \emph{{Non-singular string cosmology via
  $\alpha^{\prime}$ corrections}},
  \href{https://doi.org/10.1007/JHEP10(2019)263}{\emph{JHEP} {\bfseries 10}
  (2019) 263} [\href{https://arxiv.org/abs/1909.00830}{{\ttfamily
  arXiv:1909.00830}}].

\bibitem{Wang:2019dcj}
P.~Wang, H.~Wu, H.~Yang and S.~Ying, \emph{{Construct $\alpha^{\prime}$
  corrected or loop corrected solutions without curvature singularities}},
  \href{https://doi.org/10.1007/JHEP01(2020)164}{\emph{JHEP} {\bfseries 01}
  (2020) 164} [\href{https://arxiv.org/abs/1910.05808}{{\ttfamily
  arXiv:1910.05808}}].

\bibitem{Bernardo:2020zlc}
H.~Bernardo and G.~Franzmann, \emph{{$\alpha'$-Cosmology: solutions and
  stability analysis}},
  \href{https://doi.org/10.1007/JHEP05(2020)073}{\emph{JHEP} {\bfseries 05}
  (2020) 073} [\href{https://arxiv.org/abs/2002.09856}{{\ttfamily
  arXiv:2002.09856}}].

\bibitem{Bernardo:2020nol}
H.~Bernardo, R.~Brandenberger and G.~Franzmann, \emph{{String cosmology
  backgrounds from classical string geometry}},
  \href{https://doi.org/10.1103/PhysRevD.103.043540}{\emph{Phys. Rev. D}
  {\bfseries 103} (2021) 043540}
  [\href{https://arxiv.org/abs/2005.08324}{{\ttfamily arXiv:2005.08324}}].

\bibitem{Bernardo:2020bpa}
H.~Bernardo, R.~Brandenberger and G.~Franzmann, \emph{{Solution of the Size and
  Horizon Problems from Classical String Geometry}},
  \href{https://doi.org/10.1007/JHEP10(2020)155}{\emph{JHEP} {\bfseries 10}
  (2020) 155} [\href{https://arxiv.org/abs/2007.14096}{{\ttfamily
  arXiv:2007.14096}}].

\bibitem{Quintin:2021eup}
J.~Quintin, H.~Bernardo and G.~Franzmann, \emph{{Cosmology at the top of the
  $\alpha'$ tower}},  \href{https://arxiv.org/abs/2105.01083}{{\ttfamily
  arXiv:2105.01083}}.

\bibitem{Nunez:2020hxx}
C.A.~N\'u\~nez and F.E.~Rost, \emph{{New non-perturbative de Sitter vacua in
  $\alpha'$-complete cosmology}},
  \href{https://arxiv.org/abs/2011.10091}{{\ttfamily arXiv:2011.10091}}.

\bibitem{Brandenberger:1988aj}
R.H.~Brandenberger and C.~Vafa, \emph{{Superstrings in the Early Universe}},
  \href{https://doi.org/10.1016/0550-3213(89)90037-0}{\emph{Nucl. Phys. B}
  {\bfseries 316} (1989) 391}.

\bibitem{Gasperini:1991ak}
M.~Gasperini and G.~Veneziano, \emph{{O(d,d) covariant string cosmology}},
  \href{https://doi.org/10.1016/0370-2693(92)90744-O}{\emph{Phys. Lett. B}
  {\bfseries 277} (1992) 256}
  [\href{https://arxiv.org/abs/hep-th/9112044}{{\ttfamily hep-th/9112044}}].

\bibitem{Copeland:1994km}
E.J.~Copeland, A.~Lahiri and D.~Wands, \emph{{String cosmology with a time
  dependent antisymmetric tensor potential}},
  \href{https://doi.org/10.1103/PhysRevD.51.1569}{\emph{Phys. Rev. D}
  {\bfseries 51} (1995) 1569}
  [\href{https://arxiv.org/abs/hep-th/9410136}{{\ttfamily hep-th/9410136}}].

\bibitem{Angus:2018mep}
S.~Angus, K.~Cho and J.H.~Park, \emph{{Einstein Double Field Equations}},
  \href{https://doi.org/10.1140/epjc/s10052-018-5982-y}{\emph{Eur. Phys. J. C}
  {\bfseries 78} (2018) 500}
  [\href{https://arxiv.org/abs/1804.00964}{{\ttfamily arXiv:1804.00964}}].

\bibitem{Angus:2019bqs}
S.~Angus, K.~Cho, G.~Franzmann, S.~Mukohyama and J.H.~Park, \emph{{$\mathbf
  {O}(D,D)$ completion of the Friedmann equations}},
  \href{https://doi.org/10.1140/epjc/s10052-020-8379-7}{\emph{Eur. Phys. J. C}
  {\bfseries 80} (2020) 830}
  [\href{https://arxiv.org/abs/1905.03620}{{\ttfamily arXiv:1905.03620}}].

\bibitem{Battefeld:2005av}
T.~Battefeld and S.~Watson, \emph{{String gas cosmology}},
  \href{https://doi.org/10.1103/RevModPhys.78.435}{\emph{Rev. Mod. Phys.}
  {\bfseries 78} (2006) 435}
  [\href{https://arxiv.org/abs/hep-th/0510022}{{\ttfamily hep-th/0510022}}].

\bibitem{Basile:2021amb}
I.~Basile and A.~Platania, \emph{{Cosmological $\alpha'$-corrections from the
  functional renormalization group}},
  \href{https://arxiv.org/abs/2101.02226}{{\ttfamily arXiv:2101.02226}}.

\bibitem{Bonezzi:2021mub}
R.~Bonezzi, T.~Codina and O.~Hohm, \emph{{Beta functions for the
  duality-invariant sigma model}},
  \href{https://arxiv.org/abs/2103.15931}{{\ttfamily arXiv:2103.15931}}.

\bibitem{Alexandre:2006xh}
J.~Alexandre, J.R.~Ellis and N.E.~Mavromatos, \emph{{Non-Perturbative
  Formulation of Time-Dependent String Solutions}},
  \href{https://doi.org/10.1088/1126-6708/2006/12/071}{\emph{JHEP} {\bfseries
  12} (2006) 071} [\href{https://arxiv.org/abs/hep-th/0610072}{{\ttfamily
  hep-th/0610072}}].

\bibitem{Alexandre:2007dp}
J.~Alexandre and N.E.~Mavromatos, \emph{{Can strings live in four
  dimensions?}},  \href{https://arxiv.org/abs/hep-th/0703171}{{\ttfamily
  hep-th/0703171}}.

\bibitem{Alexandre:2007hs}
J.~Alexandre, N.E.~Mavromatos and D.~Tanner, \emph{{Antisymmetric-Tensor and
  Electromagnetic effects in an alpha-prime-non-perturbative Four-Dimensional
  String Cosmology}},
  \href{https://doi.org/10.1088/1367-2630/10/3/033033}{\emph{New J. Phys.}
  {\bfseries 10} (2008) 033033}
  [\href{https://arxiv.org/abs/0708.1154}{{\ttfamily arXiv:0708.1154}}].

\bibitem{Dasgupta:1999ss}
K.~Dasgupta, G.~Rajesh and S.~Sethi, \emph{{M theory, orientifolds and G -
  flux}}, \href{https://doi.org/10.1088/1126-6708/1999/08/023}{\emph{JHEP}
  {\bfseries 08} (1999) 023}
  [\href{https://arxiv.org/abs/hep-th/9908088}{{\ttfamily hep-th/9908088}}].

\bibitem{Gukov:1999ya}
S.~Gukov, C.~Vafa and E.~Witten, \emph{{CFT's from Calabi-Yau four folds}},
  \href{https://doi.org/10.1016/S0550-3213(00)00373-4}{\emph{Nucl. Phys. B}
  {\bfseries 584} (2000) 69}
  [\href{https://arxiv.org/abs/hep-th/9906070}{{\ttfamily hep-th/9906070}}].

\bibitem{Giddings:2001yu}
S.B.~Giddings, S.~Kachru and J.~Polchinski, \emph{{Hierarchies from fluxes in
  string compactifications}},
  \href{https://doi.org/10.1103/PhysRevD.66.106006}{\emph{Phys. Rev. D}
  {\bfseries 66} (2002) 106006}
  [\href{https://arxiv.org/abs/hep-th/0105097}{{\ttfamily hep-th/0105097}}].

\bibitem{Grana:2005jc}
M.~Grana, \emph{{Flux compactifications in string theory: A Comprehensive
  review}}, \href{https://doi.org/10.1016/j.physrep.2005.10.008}{\emph{Phys.
  Rept.} {\bfseries 423} (2006) 91}
  [\href{https://arxiv.org/abs/hep-th/0509003}{{\ttfamily hep-th/0509003}}].

\bibitem{Blumenhagen:2006ci}
R.~Blumenhagen, B.~Kors, D.~Lust and S.~Stieberger, \emph{{Four-dimensional
  String Compactifications with D-Branes, Orientifolds and Fluxes}},
  \href{https://doi.org/10.1016/j.physrep.2007.04.003}{\emph{Phys. Rept.}
  {\bfseries 445} (2007) 1}
  [\href{https://arxiv.org/abs/hep-th/0610327}{{\ttfamily hep-th/0610327}}].

\bibitem{Kachru:2003aw}
S.~Kachru, R.~Kallosh, A.D.~Linde and S.P.~Trivedi, \emph{{De Sitter vacua in
  string theory}},
  \href{https://doi.org/10.1103/PhysRevD.68.046005}{\emph{Phys. Rev. D}
  {\bfseries 68} (2003) 046005}
  [\href{https://arxiv.org/abs/hep-th/0301240}{{\ttfamily hep-th/0301240}}].

\bibitem{Balasubramanian:2005zx}
V.~Balasubramanian, P.~Berglund, J.P.~Conlon and F.~Quevedo, \emph{{Systematics
  of moduli stabilisation in Calabi-Yau flux compactifications}},
  \href{https://doi.org/10.1088/1126-6708/2005/03/007}{\emph{JHEP} {\bfseries
  03} (2005) 007} [\href{https://arxiv.org/abs/hep-th/0502058}{{\ttfamily
  hep-th/0502058}}].

\bibitem{Dasgupta:2019gcd}
K.~Dasgupta, M.~Emelin, M.M.~Faruk and R.~Tatar, \emph{{de Sitter Vacua in the
  String Landscape}},  \href{https://arxiv.org/abs/1908.05288}{{\ttfamily
  arXiv:1908.05288}}.

\bibitem{Brahma:2020tak}
S.~Brahma, K.~Dasgupta and R.~Tatar, \emph{{de Sitter Space as a
  Glauber-Sudarshan State}},
  \href{https://doi.org/10.1007/JHEP02(2021)104}{\emph{JHEP} {\bfseries 02}
  (2021) 104} [\href{https://arxiv.org/abs/2007.11611}{{\ttfamily
  arXiv:2007.11611}}].

\end{thebibliography}\endgroup

\end{document}